\documentclass[aps,preprintnumbers,showpacs,nofootinbib,superscriptaddress,floatfix]{revtex4}

\pdfoutput=1 % if your are submitting a pdflatex (i.e. if you have
\usepackage{graphicx}
\usepackage{amssymb}
\usepackage{amsmath}
\usepackage{bm}
\usepackage{slashed}
\usepackage{datetime}
\usepackage{mciteplus}
\usepackage{multirow}
\usepackage{color}
\usepackage[utf8]{inputenc}
\usepackage[normalem]{ulem}
\def\nslash{n\!\!\!\slash}
\def\Rslash{R\!\!\!\slash}

\newcommand{\tr}{\operatorname*{Tr}\nolimits} % Trace operator
\def\g{\gamma}

\begin{document}
\allowdisplaybreaks[2]

\title{Analogies between hadron-in-jet \\ and dihadron fragmentation}

\author{Alessandro Bacchetta}
\email{alessandro.bacchetta@unipv.it}
\affiliation{Dipartimento di Fisica, Universit\`a di Pavia, via Bassi 6,
  I-27100 Pavia} 
\affiliation{INFN Sezione di Pavia, via Bassi 6, I-27100 Pavia, Italy}

\author{Marco Radici}
\email{marco.radici@pv.infn.it}
\affiliation{INFN Sezione di Pavia, via Bassi 6, I-27100 Pavia, Italy}

\author{Lorenzo Rossi}
\email{lorenzo.rossi@pv.infn.it}
\affiliation{Dipartimento di Fisica, Universit\`a di Pavia, via Bassi 6,
  I-27100 Pavia} 
\affiliation{INFN Sezione di Pavia, via Bassi 6, I-27100 Pavia, Italy}

\begin{abstract}
We describe the formal analogies in the description of the inclusive production in hard processes of hadron pairs (based on dihadron fragmentation functions) and of a single hadron inside a jet (based on hadron-in-jet fragmentation functions). Since several observables involving dihadron fragmentation functions have been proposed in the past, we are able to suggest new interesting observables involving hadron-in-jet fragmentation functions, in lepton-hadron deep-inelastic scattering and hadronic collisions. 
\end{abstract}

%\date{\today, \currenttime}

%\pacs{ }

\maketitle
%\tableofcontents

%%%%%%%%%%%%%%%%%%%%%%%%%%%%%%%%%%%%%%%%%%%%%%%%%%%%%%%%%%%%%%%%%%
\section{Introduction}
\label{s:intro}

Investigation of the partonic structure of hadrons is based on the crucial method of factorization, which makes it possible to split the cross section of a given process in a perturbative calculable hard cross section (describing the underlying elementary process at the partonic level) and one or more nonperturbative functions (describing the distribution of partons inside hadrons and/or their fragmentation into detected hadronic final states). Although factorization has been established for many hard processes in the collinear framework, where transverse momenta of all partons are integrated, this is not the case for transverse-momentum dependent partonic functions (TMDs). 

For certain processes involving two hadrons in the initial state and with observed hadronic final states, e.g., inclusive production of hadrons in hadronic collisions $A + B \to C +D  + X$, TMD factorization can be explicitly broken because the strongly interacting particles are entangled by a complicate color flow~\cite{Rogers:2010dm, Collins:2007nk}. Because of this, it is not possible to describe these processes in terms of the TMDs that appear in other processes, 
like the inclusive production of a hadron $C$ in Deep-Inelastic Scattering (Semi-Inclusive DIS -- SIDIS -- denoted as $\ell + A \to \ell' + C + X$)~\cite{Collins:2004nx,Ji:2004wu} or the inclusive production of two hadrons $C, \, D$ in electron-positron annihilations ($e^+ + e^- \to C + D + X$)~\cite{Collins:1981uk} or the Drell--Yan process~\cite{Collins:1988gx}.
Even neglecting factorization-breaking contributions, the TMDs involved in hadron-hadron collisions would be different from the ones in the other processes, an effect that has been referred to as generalized universality~\cite{Bomhof:2004aw,Bacchetta:2005rm,Buffing:2012sz,Buffing:2013kca}.

The most familiar example where this problem occurs is the study of the Collins effect~\cite{Collins:1993kk}. The so-called ``Collins function" can be used as an analyzer of the transverse polarization of the fragmenting quark. It can appear in SIDIS in combination with the chiral-odd TMD parton distribution function (TMD PDF) $h_1$, called ``transversity", and in the $e^+ + e^- \to C + D + X$ process~\cite{Anselmino:2007fs,Anselmino:2013vqa,Anselmino:2015sxa}. However, because of TMD factorization breaking, it is not possible to rigorously study the Collins function in hadronic collisions.

%In SIDIS with a transversely polarized hadronic target $A^\uparrow$, the azimuthal distribution of the final hadron $h$ is distorted by a term of the cross section proportional to $h_1 \otimes H_1^\perp$~\cite{Anselmino}: the TMD parton distribution function (TMD PDF) $h_1$ describes the probability density of finding transversely polarized quarks inside $A^\uparrow$, while the TMD fragmentation function (TMD FF) $H_1^\perp$ describes the fragmentation of a transversely polarized quark into the detected hadron $h$; the symbol $\otimes$ indicates a convolution upon the transverse momenta of involved quarks. The Collins function $H_1^\perp$ can be independently extracted by looking at the azimuthally asymmetric term $H_1^{\perp \, q} \otimes H_1^{\perp \, \bar{q}}$ in the  $e^- + e^+ \to h_1 + h_2 + X$ cross section~\cite{Anselmino}. By simultaneously analyzing the two processes, the two unknown $h_1$ and $H_1^\perp$ can be extracted from data~\cite{}. In order to confirm their universality, the Collins effect should be cross-checked in processes like $A^\uparrow + B \to h + X$, but no rigorous factorization of a $h_1 \otimes f_1 \otimes H_1^\perp$ term exists in the TMD framework (with $f_1$ the unpolarized TMD PDF). 

An alternative option to the Collins effect is represented by the inclusive production of two hadrons coming from the fragmentation of a single parton. In this case, the analyzer of the transverse polarization of the fragmenting quark is represented by the transverse component of the relative momentum of the hadron pair~\cite{Collins:1994kq}. The advantage is that this correlation survives the integration over parton transverse momenta and can be analyzed in the collinear framework. Hence, in the SIDIS process $\ell + A^\uparrow \to \ell' + (C_1 \, C_2) + X$ the transversity $h_1$ can be extracted as a collinear PDF through the chiral-odd partner $H_1^\sphericalangle$, the dihadron fragmentation function (DiFF) that describes the fragmentation of a transversely polarized quark into the hadron pair~\cite{Jaffe:1998hf,Radici:2001na,Bacchetta:2002ux} and is also called Interference Fragmentation Function (IFF)~\cite{Jaffe:1998hf,Radici:2001na}. As for the Collins function, the $H_1^\sphericalangle$ can be independently extracted from azimuthal asymmetries in the production of two opposite dihadron pairs in $e^+ e^-$ annihilations, in the collinear framework~\cite{Boer:2003ya,Matevosyan:2018icf,Courtoy:2012ry,Radici:2015mwa}. This last remark makes the crucial difference. First of all,  it allows to cross-check the universality of both $h_1$ and $H_1^\sphericalangle$ in hadronic collisions of the type $A + B^\uparrow \to (C_1 \, C_2) + X$~\cite{Radici:2016lam}. Secondly, it makes it possible to extract the chiral-odd PDF $h_1$ from a global fit of SIDIS, $e^+ e^-$ and hadronic collision data in the same theoretically rigorous way as it is usually done for the other unpolarized $f_1$ and helicity $g_1$ PDFs~\cite{Radici:2018iag}. 

Another intriguing option is represented by the inclusive production of a hadron inside a jet. In fact, for a collision process like $A + B \to (\mathrm{Jet} \, C) + X$ the cross section can be factorized in a hybrid form~\cite{Yuan:2007nd}: it involves collinear PDFs in the initial collision, but the final state is represented by a new function, the jet TMDFF (jTMDFF)~\footnote{In Ref.~\cite{Kang:2017glf}, the jTMDFF is called semi-inclusive TMD fragmenting jet function (siTMDFJF).}, that depends on the jet kinematics. The jTMDFF can be matched onto the same TMD FF of hadron $C$ which appears in SIDIS and $e^+ e^-$ cross sections in the TMD framework~\cite{Kang:2017glf}. It is then possible to access TMD FFs even for that class of processes where factorization in the TMD framework is not available. When one of the two colliding hadrons is transversely factorized, say $B^\uparrow$, the fragmentation of the transversely polarized quark is described by the  polarized jTMDFF ${\cal H}_1^\perp$ that can be matched onto the Collins function $H_1^\perp$~\cite{Kang:2017btw}: this ``Collins-in-jet" effect makes it possible to check the universality of the Collins function and gives an alternative option to access the transversity $h_1$ in a rigorously factorized framework. 

The hybrid factorization for the hadron-in-jet inclusive production has been shown to work also for the SIDIS cross section~\cite{Liu:2018trl,Arratia:2020nxw,Kang:2021ffh}. Hence, it comes natural to consider the formal similarities between the inclusive production of dihadrons and of hadrons inside a jet, {\it i.e.} between DiFFs and jTMDFFs. In this way, we are able to transfer the knowledge acquired on one mechanism to the other one, and suggest new channels to investigate the partonic structure of hadrons. 

The paper is organized as follows. In Sec.~\ref{s:DiFF}, we recall the formalism for describing the inclusive dihadron production in unpolarized proton-proton collisions. In Sec.~\ref{s:jTMDFF}, we illustrate the formulae for the inclusive hadron-in-jet production in the same process. In Sec.~\ref{s:chiralodd}, we generalize the formalism to the case of collisions with one transversely polarized hadron. In Sec.~\ref{s:pp2h_h-in-jet}, by comparing the cross sections for the two mechanisms we establish a general set of correspondence rules. In Sec.~\ref{s:chances}, we use these rules to extend the study of two processes: a) the inclusive production of two back-to-back hadrons-in-jet in unpolarized proton-proton collisions, which could give access to jets initiated by linearly polarized gluons; b) the inclusive production of a hadron-in-jet in SIDIS up to subleading twist, which could give access to the chiral-odd PDF $e(x)$ related to the nucleon scalar charge. Finally, in Sec.~\ref{s:end} we conclude and give some future perspectives.

%%%%%%%%%%%%%%%%%%%%%%%%%%%%%%%%%%%%%%%%%%%%%%%%%%%%%%%%%%%%%%%%%%

%%%%%%%%%% Fig. 1  %%%%%%%%%
\begin{figure}[ht]
\begin{center}
\includegraphics[width=0.7\textwidth]{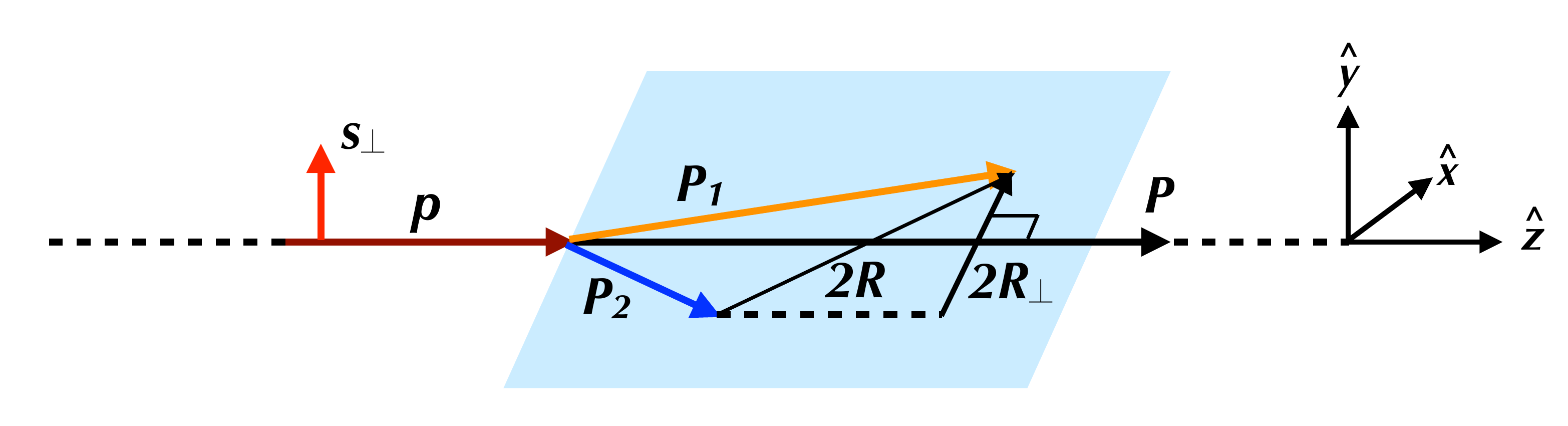} 
\end{center}
\caption{%\textcolor{red}{Ale: The transverse momenta $P_{1\perp}$ and $P_{2\perp}$ are pointing in the $y$ direction? Shouldn't they point in a generic transverse direction? The vector $R_{\perp}$ should also be indicated.} 
Collinear kinematics for the fragmentation of a quark with 3-momentum $\bm{p}$ (and transverse polarization $\bm{s}_\perp$) into a pair of hadrons with total 3-momentum $\bm{P} = \bm{P}_1 + \bm{P}_2$ pointing along $\bm{p}$, namely with $\bm{P}_\perp = 0$. The $\hat{z}$ axis is along the same direction of $\bm{p}$ and $\bm{P}$. }
%The transverse components of vectors with respect to the $\hat{z}$ axis are denoted with subscript $_\perp$.
\label{f:2h_kin}
\end{figure}
%%%%%%%%%%%%%%%%%%%%%%%

%%%%%%%%%%%%%%%%%%%%%%%%
\section{Fragmentation into a pair of hadrons}
\label{s:DiFF}

We consider the fragmentation of an unpolarized quark $q$, with 4-momentum $p$ and mass $m$, into two unpolarized hadrons inside the same jet, with 4-momenta $P_1, \  P_2$ and masses $M_1, \  M_2$, respectively. We define the total 4-momentum $P = P_1 + P_2$ and relative 4-momentum $R = (P_1 - P_2)/2$ of the pair, where $P^2 = M_{hh}^2$ is its invariant mass. We choose the $\hat{z}$ axis along the direction of the jet axis. At leading order in the strong coupling constant (LO), we  identify the jet axis with the direction of the 3-momentum $\bm{p}$. We choose the so-called ``collinear" kinematics where the 3-momentum $\bm{P}$ is pointing along $\bm{p}$. The transverse components of $R$ with respect to $\hat{z}$ is denoted by $R_\perp$, with $R_\perp^2 = - \bm{R}_\perp^2$ (see Fig.~\ref{f:2h_kin}).~\footnote{In the following, we adhere the conventions adopted in Ref.~\cite{Boer:2011fh}: the fragmenting quark momentum is denoted by $p$, the transverse component of hadron and quark vectors with respect to each other is denoted by the $_{\perp}$ subscript, 
%and the measurable transverse component of vectors is denoted 
in all other cases with the $_{T}$ subscript.}
%\textcolor{red}{I have some difficulties with the $P_{i \perp}$ vectors: since the drawing is in collinear kinematics, it is not clear why they are not measurable (they seem to be the trans. momenta with respect to $P$) and why they are different from $R_T$. Is it necessary to use them or could we just talk about $R_T$?}

The hadron pair is inclusively produced from a hard process in deep-inelastic regime. When specifying the kinematics on the light cone, the dominant components $P_1^-, \  P_2^-, \  p^-$ can be used to define the following invariants~\cite{Bacchetta:2002ux,Bacchetta:2003vn,Courtoy:2012ry,Pisano:2015wnq}~\footnote{In Refs.~\cite{Bianconi:1999cd,Bianconi:1999uc,Radici:2001na,Boer:2003ya,Matevosyan:2018icf}, the less symmetric definition $\xi = (\zeta + 1)/2 = z_1/z = 1 - z_2/z$ was adopted.}
\begin{align}
z_{hh} &= \frac{P^-}{p^-} =  \frac{P_1^- + P_2^-}{p^-} = z_1 + z_2 \nonumber \\
\zeta &= \frac{2 R^-}{P^-} = \frac{z_1 - z_2}{z_{hh}} \; ,
\label{e:2h_momfrac}
\end{align}
which represent the fraction of the fragmenting quark momentum carried by the hadron pair and how this fraction is split inside the pair, respectively. 

The fragmentation is described starting from the quark-quark %correlator~\cite{Bianconi:1999cd,Boer:2003ya,Matevosyan:2018icf}
correlator~\cite{Bacchetta:2002ux,Bacchetta:2003vn}
\begin{equation}
\Delta (p, P, R) = \sum_X \int \frac{dx}{(2\pi)^4} \, e^{i p\cdot x} \, \langle 0 | \psi (x) | X, P, R \rangle \langle X, P, R | \bar{\psi} (0) | 0 \rangle \; , 
\label{e:2h_correl}
\end{equation}
where $\psi$ is the quark field operator and the sum runs over all possible final states $|X, P, R \rangle$ containing a  hadron pair with total and relative momenta $P, R,$ respectively. At leading twist, the fragmentation of an unpolarized quark into two unpolarized hadrons can be parametrized in terms of a single DiFF according to~\cite{Bacchetta:2003vn}
\begin{equation}
D_1 (z_{hh}, \zeta, \bm{R}_\perp^2) = 4 \pi \, \tr \left[ \Delta (z_{hh}, \zeta, \bm{R}_\perp^2) \, \g^- \right] \; ,
\label{e:2h_proj}
\end{equation}
where 
\begin{equation}
%\Delta (z, \zeta, \bm{R}_\perp^2) = \frac{1}{32 z} \int dp^+ \int d\bm{p}_\perp \, \Delta (p, P, R) \Big\vert_{p^- = P^- / z} \; .
\Delta (z_{hh}, \zeta, \bm{R}_\perp^2) = \frac{z_{hh}}{32} \int dp^+ \int d\bm{p}_\perp \, \Delta (p, P, R) \Big\vert_{p^- = P^- / z_{hh}} \; .
\label{e:coll_correl}
\end{equation}
%\cMR{it should be $z/32$ because of $z^2 \int d\bm{p}_\perp$, see Eqs.(16) and (18) of Ref.~\cite{Bacchetta:2003vn} and Eq.(49) of Ref.~\cite{Bacchetta:2002ux}, which are btw all consistent with Eq.~\eqref{e:2h_proj} above.}
In fact, the full dependence of the correlator in Eq.~\eqref{e:2h_correl} is reduced to the one in Eq.~\eqref{e:coll_correl} by considering that~\cite{Bianconi:1999cd} 
\begin{itemize}

\item[--]  in Eq.\eqref{e:coll_correl} we integrate over the light-cone suppressed variable $p^+$ and over $\bm{p}_\perp$ with the condition $p^- = P^- / z_{hh}$; 

\item[--]  our choice of frame and kinematics implies no dependence on $\bm{P}_\perp$;

\item[--] the following kinematical relations hold~\cite{Bacchetta:2002ux}:
\begin{align}
P^2 &= M_{hh}^2 \; , \quad R^2 = \frac{M_1^2 + M_2^2}{2} - \frac{M_{hh}^2}{4} - \frac{(M_1^2 - M_2^2)^2}{4 M_{hh}^2} \; , \nonumber \\
\bm{R}_\perp^2 &=\frac{1}{2} \left[ \frac{(1-\zeta) \, (1+\zeta)}{2} \, M_{hh}^2 - (1- \zeta) M_1^2 - (1+\zeta) M_2^2 \right] \; . 
\label{e:2h_invar}
\end{align}
%with $\zeta = 2 R^- / P^-$ defined in Eq.~\eqref{e:2h_momfrac}. 

\end{itemize}

It is useful to recall also that~\cite{Bacchetta:2002ux}
\begin{equation}
p\cdot R = \frac{M_1^2 - M_2^2 - \frac{\zeta}{2} M_{hh}^2}{2 z_{hh}} + z_{hh} \zeta \frac{p^2 + \bm{p}_\perp^2}{2} - \bm{p}_\perp \cdot \bm{R}_\perp \; ,
\end{equation}
from which we deduce that in general DiFFs depend only on the relative angle between $\bm{p}_\perp$ and $\bm{R}_\perp$.

%%%%%%%%%% Fig. 2  %%%%%%%%%
\begin{figure}[ht]
\begin{center}
\includegraphics[width=0.48\textwidth]{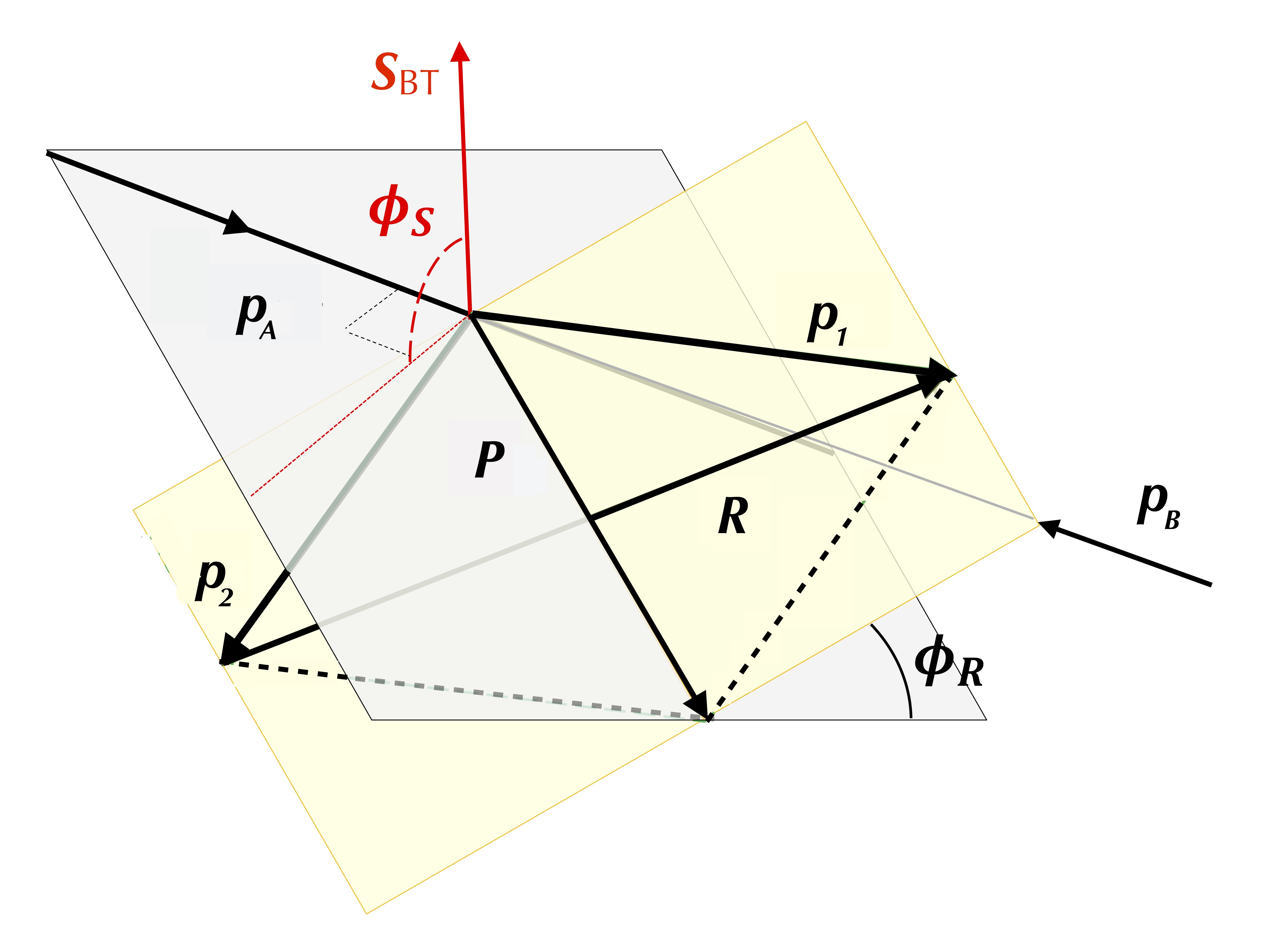} 
\end{center}
\caption{Kinematics for the collision of a proton with 3-momentum $\bm{P}_A$ and a (transversely polarized) proton with momentum $\bm{P}_B$ (and polarization $\bm{S}_{BT}$), inclusively producing two unpolarized hadrons with total and relative momenta $\bm{P} = \bm{P}_1 + \bm{P}_2$ and $\bm{R} = ( \bm{P}_1 - \bm{P}_2 )/2$. The plane formed by $\bm{P}$ and $\bm{R}$ is oriented by the azimuthal angle $\phi_R$ around $\bm{P}$ with respect to the reaction plane formed by $\bm{P}_A$ and $\bm{P}$.}
\label{f:pp2h_kin}
\end{figure}
%%%%%%%%%%%%%%%%%%%%%%%

%%%%%%%%
\subsection{Cross section for dihadron production in proton--proton collisions}
\label{s:pp2h}

If the hadron pair is inclusively produced from the collision of two unpolarized protons with momenta $P_A$ and $P_B$, we can identify the reaction plane as the plane formed by $\bm{P}_A$ and $\bm{P}$. The azimuthal orientation around $\bm{P}$ of the plane formed by $\bm{P}_1$ and $\bm{P}_2$ with respect to the reaction plane is described by the azimuthal angle $\phi_R$ (see Fig~\ref{f:pp2h_kin} and Ref.~\cite{Bacchetta:2004it} for a formal definition). The transverse component of $\bm{P}$ with respect to $\bm{P}_A$ is denoted by $\bm{P}_T$. Its modulus represents the hard scale of the process, namely we assume that $|\bm{P}_T| \gg M_{hh}, \, M_1, \, M_2$. For simplicity, in the following the dependence of DiFFs on $|\bm{P}_T|$ is understood. 

At leading order in $1/|\bm{P}_T|$, the differential cross section for the process $A + B \to (C_1 \, C_2) + X$ reads (see App.~\ref{a:cross-check} and Eq.~(15) of Ref.~\cite{Bacchetta:2004it})
\begin{align}
\frac{d\sigma_{UU}}{d\eta \, d|\bm{P}_T| \, d\zeta \, d\bm{R}_\perp} &=  \sum_{a,b,c,d} \int \frac{dx_A dx_B dz_{hhC}}{x_A x_B z_{hhC}^2} \, f_1^a (x_A)\, f_1^b (x_B) \, 
\frac{|\bm{P}_T|\, \hat{s}}{2\pi s} \, \frac{d\hat{\sigma}_{ab\to cd}}{d\hat{t}} \, \hat{s} \delta(\hat{s} + \hat{t} + \hat{u})  \, D_1^c (z_{hhC}, \zeta, \bm{R}_\perp^2)  \; , 
\label{e:pp2h_xsec}
\end{align}
where $f_1^a$ and $f_1^b$ are the usual parton distribution functions (PDFs) in the proton for partons $a, \, b$ with fractional momenta $x_A, \, x_B,$ respectively, and $\eta$ is the pseudorapidity of the hadron pair with respect to $\bm{P}_A$:
\begin{equation}
\eta = \frac{1}{2} \, \log \frac{P^0 + P_z}{P^0 - P_z} \; .
\label{e:eta}
\end{equation}
The elementary cross section $d\hat{\sigma}$ describes the scattering of partons $a$ and $b$ into partons $c$ (with momentum $P / z_{hhC}$) and $d$, which is not detected.
%where the former has momentum $P / z_{hhC}$ and we inclusively sum upon the latter. 
The partonic Mandelstam variables $\hat{s}, \, \hat{t}, \, \hat{u}$ are related to the external ones by
\begin{equation}
\hat{s} = x_A \, x_B \, s \; , \quad \hat{t} = \frac{x_A}{z_{hhC}}\, t \; , \quad \hat{u} = \frac{x_B}{z_{hhC}}\, u \; .
\label{e:hatMandelst}
\end{equation}
The $\delta$ function in Eq.~\eqref{e:pp2h_xsec} expresses the momentum conservation in the partonic scattering, and it can be rewritten as~\cite{Bacchetta:2004it}
\begin{equation}
\hat{s} \, \delta (\hat{s} + \hat{t} + \hat{u})  = z_{hhC} \, \delta (z_{hhC} - \bar{z}_{hh}) \; ,
\label{e:ab2cd_mom}
\end{equation}
where 
\begin{equation}
\bar{z}_{hh} = \frac{|\bm{P}_T|}{\sqrt{s}} \, \frac{x_A \, e^{-\eta} + x_B \, e^{\eta}}{x_A \, x_B} \; .
\label{e:zbar}
\end{equation}
In Eq.~\eqref{e:pp2h_xsec}, the sum runs upon all possible combinations of parton flavors. The elementary cross sections $d\hat{\sigma}_{ab\to cd}$ for the independent combinations are listed in the Appendix of Ref.~\cite{Bacchetta:2004it}.

%%%%%%%%%%%%%%%%%%%%%%%%%%%%%%%%%%%%%%%%%%%%%%%%%%%%%%%%%%%%%%%%%%

%%%%%%%%%%%%%%%%%%%%%%%
\section{Hadron-in-jet fragmentation}
\label{s:jTMDFF}

\begin{figure}[ht]
\begin{center}
\includegraphics[width=0.7\textwidth]{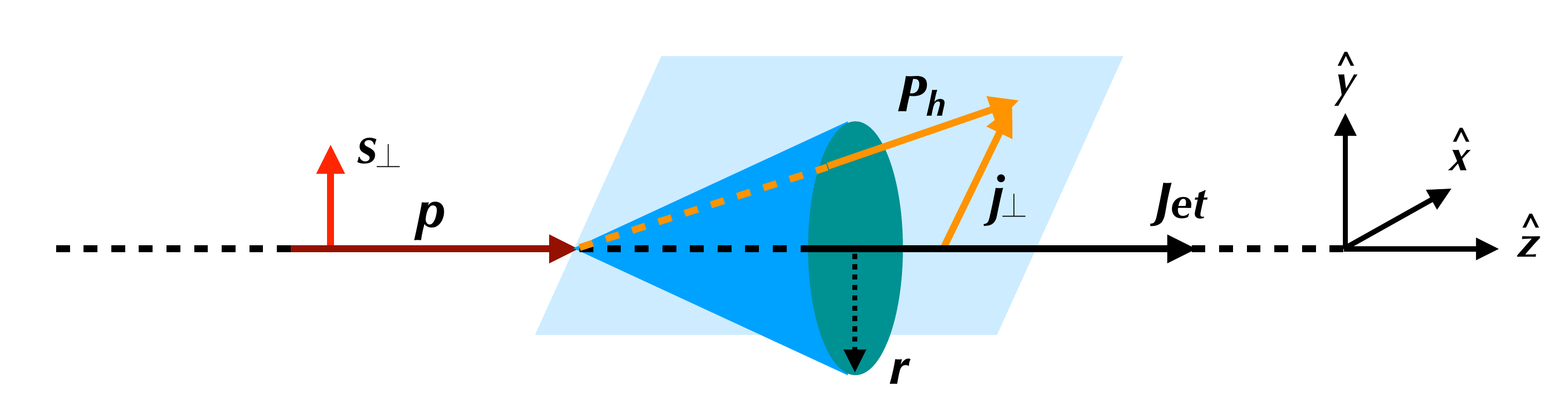} 
\end{center}
\caption{%\textcolor{red}{Ale:The transverse momentum $j_{\perp}$ points in the $y$ direction? Shouldn't it point in a generic transverse direction?}
Fragmentation of a unpolarized hadron with 3-momentum $\bm{P}_h$ inside a jet of radius $r$ initiated by a quark with 3-momentum $\bm{p}$ (and transverse polarization $\bm{s}_\perp$). The transverse component of $\bm{P}_h$ with respect to the standard jet axis is denoted by $\bm{j}_\perp$~\cite{Kang:2017glf}. For sake of simplicity, $\bm{p}$ and the jet axis are approximately taken along the same direction.}
\label{f:h-in-jet_kin}
\end{figure}

We now consider the distribution of a hadron with 4-momentum $P_h$ and mass $M_h$ inside a jet with radius $r$, initiated by a unpolarized quark $q$ with 4-momentum $p$ and mass $m$. Following Ref.~\cite{Kang:2017glf}, we denote by $\bm{j}_\perp$ the transverse momentum of the hadron inside the jet (see Fig.~\ref{f:h-in-jet_kin}). The latter is defined with respect to the standard jet axis (rather than using a recoil-free algorithm) because only in this case a direct connection to the TMD FF can be made~\cite{Kang:2017glf}. As in the dihadron case, the $\hat{z}$ axis is chosen along the standard jet axis and at LO it is identified with the direction of $\bm{p}$. 

%%%%%%%%%% Fig. 3  %%%%%%%%%
%%%%%%%%%%%%%%%%%%%%%%%

When the jet is produced in a hard process in deep-inelastic kinematical regime, the large light-cone components of quark, jet, and hadron vectors are denoted by $p^-, \, J^-,$ and $P_h^-$, respectively. They are used to define the following invariants
\begin{equation}
z_J = \frac{J^-}{p^-}  \; , \qquad z_h = \frac{P_h^-}{J^-} \; ,
\label{e:h-in-jet_momfrac}
\end{equation}
which represent the fraction of the fragmenting quark momentum carried by the jet and the fraction of jet momentum carried by the hadron inside the jet, respectively. The $J^-$ is related to the transverse momentum of the reconstructed jet in the hard process, whose size is denoted as $|\bm{P}_T|$ and represents the hard scale of the process itself. 

The fragmentation is described starting from the quark-quark correlator
\begin{equation}
\Delta (p, J, P_h) = \sum_X \int \frac{dx}{(2\pi)^4} \, e^{i p\cdot x} \, \langle 0 | \psi (x) | X, J, P_h \rangle \langle X, J, P_h | \bar{\psi} (0) | 0 \rangle \; , 
\label{e:h-in-jet_correl}
\end{equation}
where, as before, $\psi$ is the quark field operator and the sum runs over all possible final states $|X, J, P_h \rangle$ containing a  hadron $P_h$ inside a jet $J$. At leading twist, the object describing the observed hadron inside the produced jet can be parametrized in terms of a jTMDFF according to~\cite{Kang:2017glf}
\begin{equation}
{\cal D}_1 (z_J, z_h, \bm{j}_\perp;  |\bm{P}_T|,  |\bm{P}_T| r) = \frac{z_J}{4 N_c} \, \tr \left[ \Delta (z_J, z_h, \bm{j}_\perp,  |\bm{P}_T|, |\bm{P}_T| r) \, \g^- \right] \; ,
\label{e:h-in-jet_proj}
\end{equation}
where $N_c$ is the number of quark colors, $|\bm{P}_T| r$ is the typical momentum scale of the jet~\cite{Kang:2017glf}, and 
\begin{equation}
\Delta (z_J, z_h, \bm{j}_\perp;  |\bm{P}_T|, |\bm{P}_T| r) = \int dp^+ \, \Delta (p, J, P_h)\Biggr|_{\scriptsize\begin{array}{lcc} p^-  = J^- / z_J \\ 
|\bm{p}_T|  =  |\bm{P}_T| / z_J \\
J^-  = P_h^- / z_h \end{array}}  \; .
\label{e:tmdff_correl}
\end{equation}

Depending on the relative size of $|\bm{j}_\perp|$, $|\bm{P}_T| r$ and the QCD nonperturbative scale $\Lambda_{\rm QCD}$, the jTMDFF of Eq.~\eqref{e:h-in-jet_proj} can be expressed in different factorized forms. Here, we are interested in the kinematical region $\Lambda_{\rm QCD} \lesssim |\bm{j}_\perp| \ll |\bm{P}_T| r$ where collinear radiation within the jet and soft radiation of order $|\bm{j}_\perp|$ are relevant, while harder radiation is allowed only outside the jet and it does not affect the distribution of the hadron transverse momentum $|\bm{j}_\perp|$. In this regime, a  factorized form for ${\cal D}_1$ is given in Ref.~\cite{Kang:2017glf} in terms of a hard matching function (related to the hard out-of-jet radiation) and a convolution of a usual TMD FF and a soft function (accounting for the soft radiation inside the jet). It is obtained by initially evolving the TMD FF in the usual Collins--Soper--Sterman (CSS) scheme up to the jet scale $|\bm{P}_T| r$, then matching to the calculable hard function describing the out-of-jet radiation, and finally evolving to the hard scale by using the standard time-like DGLAP equations. All calculations in Ref.~\cite{Kang:2017glf} are performed at NLO. At LO, the direction of the quark momentum $\bm{p}$ coincides with the standard jet axis and its transverse component is equal to the transverse momentum of the reconstructed jet in the hard process, $|\bm{p}_T| \approx |\bm{P}_T|$. In this approximation, the jTMDFF ${\cal D}_1^q$ for the fragmentation of a quark $q$ into a hadron inside the jet reduces to 
%is given in Eq.~(2.64) of Ref.\cite{Kang:2017glf}, which at LO reads 
\begin{align}
{\cal D}_1^q (z_J, z_h, \bm{j}_\perp;  |\bm{P}_T|, |\bm{P}_T| r) \Big\vert_{\rm LO} &= \sum_i \, \delta (1-z_J ) \, \delta_{q i} \, D_1^i (z_h,  \bm{j}_\perp^2;  |\bm{P}_T|) \nonumber \\
&= \delta (1-z_J) \, D_1^q (z_h,  \bm{j}_\perp^2;  |\bm{P}_T|) \; ,
\label{e:jTMDFF_LO}
\end{align}
where $D_1^q$ is the standard single-hadron TMD FF that can be isolated also in $e^+ e^-$ annihilations or in semi-inclusive deep-inelastic scattering.

%%%%%%%%%%%%%%%%%%
\subsection{Cross section for hadron-in-jet fragmentation in proton--proton collisions}
\label{s:ppJh}

We consider the same situation as in Sec.~\ref{s:pp2h}, namely the collision of two unpolarized protons with momenta $P_A$ and $P_B$. The final state is now described by the inclusive production of a jet where a hadron is identified inside it with transverse momentum $\bm{j}_\perp$ with respect to the standard jet axis. Following Ref.~\cite{Kang:2017btw}, the factorization theorem for the process $A + B \to ({\rm Jet} \, C) + X$ can be written as
\begin{align}
\frac{d\sigma_{UU}}{d\eta \, d|\bm{P}_T| \, dz_h \, d\bm{j}_\perp} &= \frac{2 \pi |\bm{P}_T|}{s}  \sum_{a,b,c,d} \int \frac{dx_A dx_B dz_{JC}}{x_A x_B z_{JC}^2} \, f_1^a (x_A)\, f_1^b (x_B) \, H^U_{ab \to cd} \, {\cal D}_1^c (z_{JC}, z_h, \bm{j}_\perp;  |\bm{P}_T|, |\bm{P}_T| r) \, z_{JC}^2 \delta(z_{JC} - \bar{z}_J) \; ,
\label{e:h-in-jet_xsec}
\end{align}
where $\bar{z}_J$ is given as in Eq.~\eqref{e:zbar} and $H^U_{ab \to cd}$ describes the elementary hard process $a+b \to c+d$ from which the parton $c$ initiates the reconstructed jet. 
%It depends on the partonic center-of-mass energy $\hat{s}$ (see Eq.~\eqref{e:hatMandelst}), transverse momentum $p_T$ (related to $P_T / z_c$, see Eq.~\eqref{e:tmdff_correl}), and pseudorapidity $\hat{\eta} = \eta - \log (x_a / x_b) / 2$. 

As detailed in App.~\ref{a:hard_xsec}, the $H^U_{ab \to cd} $ of Ref.~\cite{Kang:2017btw} can be reconnected to the $d\hat{\sigma}_{ab\to cd}$ of Eq.~\eqref{e:pp2h_xsec} by
\begin{align}
H^U_{ab \to cd} &= \frac{\hat{s}}{\pi z_{JC}}\, \frac{d\hat{\sigma}_{ab \to cd}}{d\hat{t}}  \; . 
\label{e:hard_cmp}
\end{align} 

The cross section of Eq.~\eqref{e:h-in-jet_xsec} can then be cast in the form
\begin{align}
\frac{d\sigma}{d\eta \, d|\bm{P}_T| \, dz_h \, d\bm{j}_\perp} &= \sum_{a,b,c,d} \int \frac{dx_A dx_B dz_{JC}}{x_A x_B z_{JC}^2} \, f_1^a (x_A)\, f_1^b (x_B) \, 2 \frac{|\bm{P}_T| \, \hat{s}}{s} \, \frac{d\hat{\sigma}_{ab \to cd}}{d\hat{t}} \, z_{JC} \, \delta (z_{JC} - \bar{z}_J) \, {\cal D}_1^c (z_{JC}, z_h, \bm{j}_\perp;  |\bm{P}_T|, |\bm{P}_T| r) \nonumber \\
&= \sum_{a,b,c,d} \int \frac{dx_A dx_B dz_{JC}}{x_A x_B z_{JC}^2} \, f_1^a (x_A)\, f_1^b (x_B) \, 2 \frac{|\bm{P}_T| \, \hat{s}}{s} \, \frac{d\hat{\sigma}_{ab \to cd}}{d\hat{t}} \, \hat{s} \delta(\hat{s} + \hat{t} + \hat{u}) \, {\cal D}_1^c (z_{JC}, z_h, \bm{j}_\perp;  |\bm{P}_T|, |\bm{P}_T| r) \; ,
\label{e:h-in-jet_cmp}
\end{align}
where we used Eq.~\eqref{e:ab2cd_mom} adapted to the case of hadron-in-jet fragmentation, {\it i.e.}, by replacing $z_{hhC}$ with $z_{JC}$ for the fragmenting parton $c$ and using the pseudorapidity of the jet with respect to $\bm{P}_A$.

%%%%%%%%%%%%%%%%%%%%%%%%%%%%%%%%%%%%%%%%%%%%%%%%%%%%%%%%%

\section{Fragmentation of transversely polarized quarks}
\label{s:chiralodd}

We extend our study to the case of a fragmenting quark with transverse polarization $\bm{s}_\perp$. We first consider the fragmentation into a pair of unpolarized hadrons (see Fig.~\ref{f:2h_kin}). In the kinematic conditions described in Sec.~\ref{s:DiFF}, the leading-twist correlator of Eq.~\eqref{e:coll_correl} can be expanded as~\cite{Bacchetta:2003vn}
\begin{align}
\Delta (z_{hh}, \zeta, \bm{R}_\perp^2) &= \frac{1}{16 \pi}\, \left\{ D_1 (z_{hh}, \zeta, \bm{R}_\perp^2) \, \nslash_- + \, H_1^{\sphericalangle} (z_{hh}, \zeta, \bm{R}_\perp^2) \, \frac{i}{2 M_{hh}} \, [ \Rslash_\perp , \, \nslash_- ]  \right\}  \; ,
\label{e:coll_correl2}
\end{align}
where $H_1^{\sphericalangle}$ describes the probability density for a transversely polarized quark to fragment into a pair of unpolarized hadrons with total momentum collinear with the quark momentum. The $H_1^\sphericalangle$ can be extracted by the following projection
\begin{align}
\frac{(\bm{s}_\perp \times \bm{R}_\perp )\cdot \bm{P}}{M_{hh}} \   H_1^{\sphericalangle} (z_{hh}, \zeta, \bm{R}_\perp^2) &= 4 \pi \, \tr \left[ \Delta (z_{hh}, \zeta, \bm{R}_\perp^2) \, i\, \sigma^{i -} \, \g_5 \right] \; ,
\label{e:2h_proj_pol}
\end{align}
where $\sigma^{\mu \nu} = i [\g^\mu, \, \g^\nu] / 2$ and its spatial index $i$ points in the direction of $\bm{s}_\perp$. 

Similarly, if the transversely polarized quark fragments into a hadron inside a jet in the kinematical conditions described in Sec.~\ref{s:jTMDFF} (see Fig.~\ref{f:h-in-jet_kin}), we can project out the ``Collins-in-jet" function ${\cal H}_1^\perp$ from the correlator in Eq.~\eqref{e:tmdff_correl} as
\begin{align}
\frac{(\bm{s}_\perp \times \bm{j}_\perp )\cdot \bm{p}}{z_h \, M_h} \   {\cal H}_1^\perp (z_J, z_h, \bm{j}_\perp; |\bm{P}_T|,  |\bm{P}_T| r) &= \frac{z_J}{4 N_c} \, \tr \left[ \Delta (z_J, z_h, \bm{j}_\perp; |\bm{P}_T|, |\bm{P}_T| r) \,  i\, \sigma^{i -} \, \g_5 \right] \; , 
\label{e:h-in-jet_proj_pol}
\end{align}
where again the spatial index $i$ of $\sigma^{\mu \nu}$ points in the direction of $\bm{s}_\perp$. 

In the following section, for the two fragmentation scenarios we analyze the contributions that arise in the cross section for proton-proton collisions when one of the two protons is transversely polarized.

%%%%%%%%%% Fig. 4  %%%%%%%%%
\begin{figure}[ht]
\begin{center}
\includegraphics[width=0.48\textwidth]{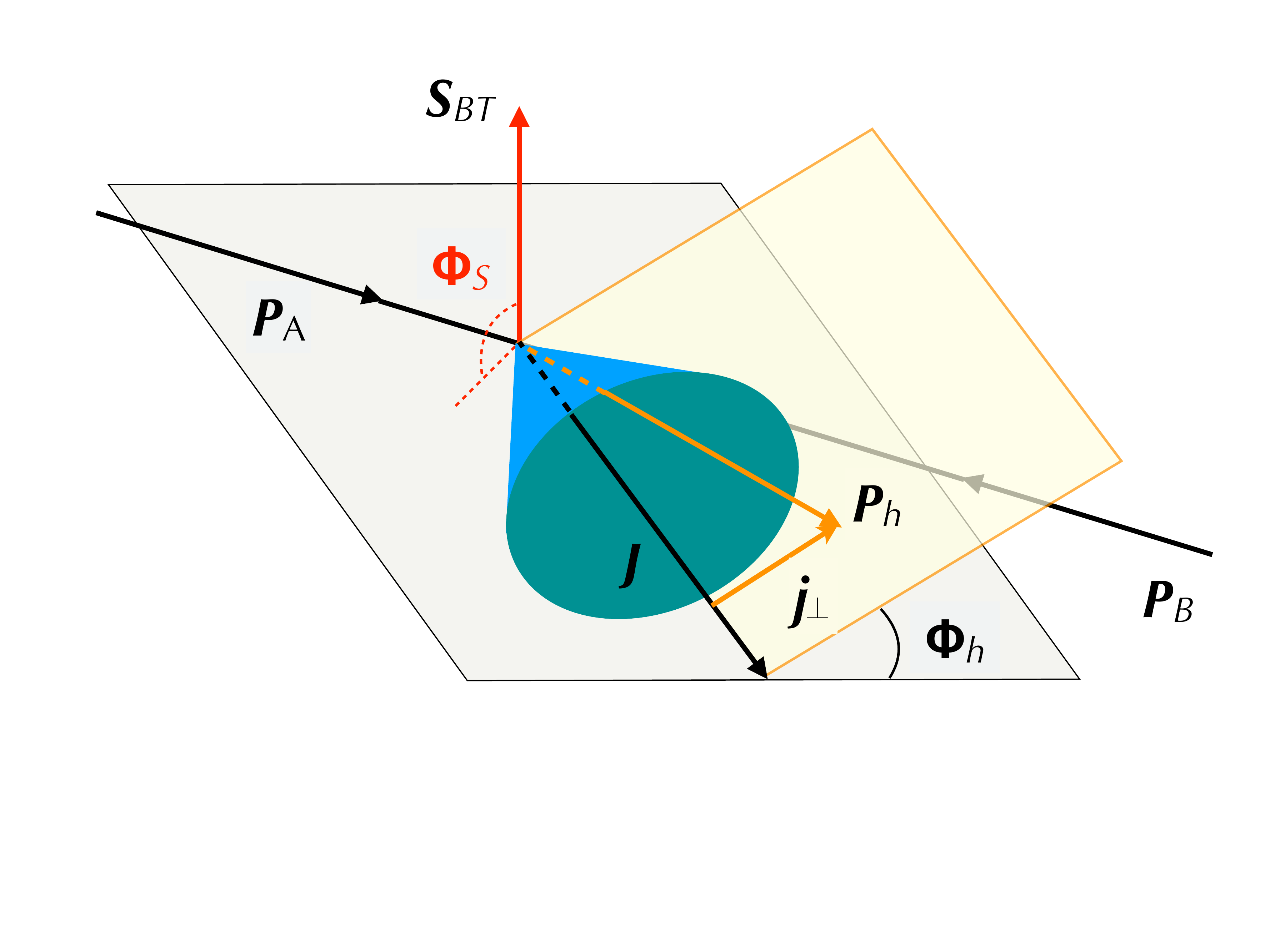} 
\end{center}
\caption{Kinematics for the collision of a proton with 3-momentum $\bm{P}_A$ and a transversely polarized proton with momentum $\bm{P}_B$ (and polarization $\bm{S}_{BT}$), inclusively producing inside a jet a hadron with 3-momentum $\bm{P}_h$ and transverse component $\bm{j}_\perp$ with respect to the jet axis $\hat{\bm{J}}$. The plane formed by $\hat{\bm{J}}$ and $\bm{P}_h$ is oriented by the azimuthal angle $\phi_h$ around $\hat{\bm{J}}$ with respect to the reaction plane formed by $\bm{P}_A$ and $\hat{\bm{J}}$.}
\label{f:pph-in-jet_kin}
\end{figure}
%%%%%%%%%%%%%%%%%%%%%%%

%%%%%%%%%%%%%
\subsection{Transversely polarized proton-proton collisions}
\label{s:chiralodd_pp}

For the process $A + B^\uparrow \to (C_1 \, C_2) + X$ depicted in Fig.~\ref{f:pp2h_kin}, the polarized part of the cross section reads (see App.~\ref{a:cross-check} and Eq.~(16) of Ref.~\cite{Bacchetta:2004it})
\begin{align}
\frac{d\sigma_{UT}}{d\eta \, d|\bm{P}_T| \, d\zeta \, d\bm{R}_\perp \, d\phi_{S_B}} &= \frac{ |\bm{S}_{BT}| }{4\pi^2}\,  \sin (\phi_{S_B} - \phi_R) \nonumber \\ 
&\hspace{-1.5cm} \times  \sum_{a,b,c,d} \int \frac{dx_A dx_B dz_{hhC}}{x_A x_B z_{hhC}^2} \, f_1^a (x_A)\, h_1^b (x_B) \, 
\frac{|\bm{P}_T|\, \hat{s}}{s} \, \frac{d\Delta \hat{\sigma}_{ab^\uparrow \to c^\uparrow d}}{d\hat{t}} \, \hat{s} \delta(\hat{s} + \hat{t} + \hat{u}) \, \frac{|\bm{R_\perp}|}{M_{hh}} \, H_1^{\sphericalangle c} (z_{hhC}, \zeta, \bm{R}_\perp^2)  \; , 
\label{e:ppT2h_xsec}
\end{align}
where $\bm{S}_{BT}$ is the transverse polarization of the colliding proton with orientation $\phi_{S_B}$ with respect to the reaction plane, and $h_1^b$ is the transversity distribution for the transversely polarized parton $b$ with fractional momentum $x_B$. The elementary cross sections $d\Delta\hat{\sigma}_{ab^\uparrow \to c^\uparrow d}$ describe the scattering of parton $a$ and $b$ with transfer of the transverse polarization of the latter to parton $c$ while summing on the undetected fragments from parton $d$. All the possible independent flavor combinations are listed in the Appendix of Ref.~\cite{Bacchetta:2004it}. 

The corresponding process $A + B^\uparrow \to ({\rm Jet} \, C) + X$ is displayed in Fig.~\ref{f:pph-in-jet_kin}. A hadron with 3-momentum $\bm{P}_h$ is inclusively produced inside a jet with standard axis $\hat{\bm{J}}$ from the collisions of a proton with 3-momentum $\bm{P}_A$ and a transversely polarized proton with 3-momentum $\bm{P}_B$ and polarization $\bm{S}_{BT}$. The azimuthal angles $\phi_{S_B}$ and $\phi_h$ describe the orientation of $\bm{S}_{BT}$ and of the plane formed by $\bm{P}_h$ and $\hat{\bm{J}}$, respectively, with respect to the reaction plane formed by $\bm{P}_A$ and $\hat{\bm{J}}$. The polarized part of the cross section reads~\cite{Kang:2017btw}~\footnote{The expression in Eq.~\eqref{e:h-in-jet_polxsec} differs from Ref.~\cite{Kang:2017btw} by a $1/z_h$ term because of a definition of the Collins function inherited from Ref.~\cite{Yuan:2007nd} which does not adhere to the Trento conventions~\cite{Bacchetta:2004jz}.}
\begin{align}
\frac{d\sigma_{UT}}{d\eta \, d|\bm{P}_T| \, dz_h \, d\bm{j}_\perp \, d\phi_{S_B}} &= |\bm{S}_{BT}| \, \sin (\phi_{S_B} - \phi_h) \, \frac{|\bm{P}_T|}{s}  \nonumber \\
&\hspace{-2cm}  \times \sum_{a,b,c,d} \int \frac{dx_A dx_B dz_{JC}}{x_A x_B z_{JC}^2} \, f_1^a (x_A)\, h_1^b (x_B) \, H^{Collins}_{ab^\uparrow \to c^\uparrow d} \, \frac{|\bm{j}_\perp|}{M_h} \, {\cal H}_1^{\perp \, c} (z_{JC}, z_h, \bm{j}_\perp;  |\bm{P}_T|, |\bm{P}_T| r) \, z_{JC}^2 \delta(z_{JC} - \bar{z}_J) \; ,
\label{e:h-in-jet_polxsec}
\end{align}
where $H^{Collins}_{ab^\uparrow \to c^\uparrow d}$ is the cross section for the transfer of transverse polarization in the elementary hard process $a+b^\uparrow \to c^\uparrow +d$, and ${\cal H}_1^{\perp \, c}$ is the polarized jTMDFF describing the hadron inside the jet produced by the transversely polarized fragmenting parton $c$. 

By extending the relation~\eqref{e:hard_cmp} to the polarized case involving $H^{Collins}_{ab^\uparrow \to c^\uparrow d}$ of Ref.~\cite{Kang:2017btw} and $d\Delta\hat{\sigma}_{ab^\uparrow \to c^\uparrow d}$ of Ref.~\cite{Bacchetta:2004it} (and exchanging $\hat{t} \leftrightarrow \hat{u}$ to account for the fact that the transversely polarized parton in Ref.~\cite{Kang:2017btw} is $a^\uparrow$ while in Ref.~\cite{Bacchetta:2004it} is $b^\uparrow$), we finally get
\begin{align}
& \frac{d\sigma_{UT}}{d\eta \, d|\bm{P}_T| \, dz_h \, d\bm{j}_\perp \, d\phi_{S_B}} =  |\bm{S}_{BT}| \, \sin (\phi_{S_B} - \phi_h) \nonumber \\
 &\quad \times \sum_{a,b,c,d} \int \frac{dx_A dx_B dz_{JC}}{x_A x_B z_{JC}^2} \, f_1^a(x_A) \, h_1^b(x_B) \, \frac{|\bm{P}_T| \hat{s}}{\pi s} \, z_{JC} \delta(z_{JC} - \bar{z}_J) \, \frac{d \Delta \hat{\sigma}_{ab^\uparrow\to c^\uparrow d}}{d \hat{t}} \, \frac{|\bm{j}_\perp|}{M_h} \, {\cal H}_{1}^{\perp \, c} (z_{JC}, z_h, \bm{j}_{\perp}; |\bm{P}_T|, |\bm{P}_T| r) \nonumber \\
 &= |\bm{S}_{BT}| \, \sin (\phi_{S_B} - \phi_h) \nonumber \\
 &\quad \times \sum_{a,b,c,d} \int \frac{dx_A dx_B dz_{JC}}{x_A x_B z_{JC}^2} \, f_1^a(x_A) \, h_1^b(x_B) \, \frac{|\bm{P}_T| \hat{s}}{\pi s} \, \frac{d \Delta \hat{\sigma}_{ab^\uparrow\to c^\uparrow d}}{d \hat{t}} \, \hat{s} \delta(\hat{s} + \hat{t} + \hat{u})  \, \frac{|\bm{j}_\perp|}{M_h} \, {\cal H}_{1}^{\perp \, c} (z_{JC}, z_h, \bm{j}_{\perp}; |\bm{P}_T|, |\bm{P}_T| r) \; .
 \label{e:h-in-jet_polcmp}
 \end{align}

%%%%%%%%%%%%%%%%%%%%%%%%%%%%%%%%%%%%%%%%%%%%%%%%%%%%%%
\section{Correspondence between dihadron and hadron-in-jet fragmentation}
\label{s:pp2h_h-in-jet}

We are now in the position to compare the cross sections for the $A + B^{(\uparrow)} \to (C_1 \, C_2) + X$ and $A + B^{(\uparrow)} \to ({\rm Jet} \, C) + X$ processes. We deduce that:
\begin{itemize}
\item[--] from Eq.~\eqref{e:h-in-jet_momfrac}, the combination $z_J z_h = P_h^- / p^-$ describes the fraction of fragmenting quark momentum carried by the hadron inside the jet; hence, it can be mapped onto 
%is the best proxy to 
$z_{hh}$ of Eq.~\eqref{e:2h_momfrac};

\item[--] by comparing the same two equations, we can map $z_h$ onto $\zeta z_{hh} = z_1 - z_2$, the relative fractional momentum carried by the hadron pair; thus, for both hadronic final states (dihadron and hadron inside jet) the light-cone kinematics can be described by a pair of invariants and we can establish a correspondence between these pairs, namely $(z_{hh}, \zeta) \leftrightarrow (z_J, z_h)$;

\item[--] along the same line, we can map the transverse momentum $\bm{j}_\perp$ of the hadron inside the jet with respect to the standard jet axis onto the transverse component $\bm{R}_\perp$ of the hadron pair relative momentum with respect to the direction of the pair total momentum, which in collinear kinematics coincides with the standard jet axis; obviously, the same mapping holds for their azimuthal angles, {\it i.e.}, $\phi_h \leftrightarrow \phi_R$;

\item[--] by directly comparing Eqs.~\eqref{e:pp2h_xsec} with \eqref{e:h-in-jet_cmp} and  Eqs.~\eqref{e:ppT2h_xsec} with \eqref{e:h-in-jet_polcmp}, the %(un)polarized 
jTMDFF can be mapped onto the corresponding DiFF according to 
\begin{align}
4 \pi \, {\cal D}_1^q (z_J, z_h, \bm{j}_\perp; |\bm{P}_T|, |\bm{P}_T| r)  &\longleftrightarrow \, D_1^q (z_{hh}, \zeta, \bm{R}_\perp^2; |\bm{P}_T|)  
\label{e:jTMDFF_DiFF}
\\
4 \pi \, {\cal H}_1^{\perp \, q} (z_J, z_h, \bm{j}_\perp; |\bm{P}_T|, |\bm{P}_T| r)  &\longleftrightarrow \, H_1^{\sphericalangle \, q} (z_{hh}, \zeta, \bm{R}_\perp^2; |\bm{P}_T|) \; .
\label{e:jTMDFF_DiFF_pol}
\end{align}
\end{itemize}

As a final remark,  DiFFs have been extracted so far only through a LO analysis of inclusive dihadron production in $e^+ e^-$ annihilation~\cite{Courtoy:2012ry}, in combined $e^+ e^-$ and SIDIS processes~\cite{Radici:2015mwa}, and in a global fit of $e^+ e^-$, SIDIS and hadron-hadron collision data~\cite{Radici:2018iag}. The formalism of jTMDFFs is available instead up to NLO, but only in the unpolarized case~\cite{Kang:2017glf,Kang:2017btw}. However, NLO corrections separately affect the elementary hard cross section for a $2\to 2$ partonic process~\cite{Aversa:1988vb} and the OPE-expanded expression of ${\cal D}_1$~\cite{Kang:2017glf}. Hence, we can argue that the same structure of the leading-twist cross sections for inclusive dihadron production in Eqs.~\eqref{e:pp2h_xsec}, \eqref{e:ppT2h_xsec} holds also at NLO.
%, such that the above comparison between the formalisms of hadron-in-jet and dihadron fragmentations is fully consistent. 

The correspondence expressed in Eqs.~\eqref{e:jTMDFF_DiFF} and \eqref{e:jTMDFF_DiFF_pol}
represents the main result of this paper. It has been derived by considering the case of proton-proton collisions but it can be extended to all hard processes where collinear factorization holds, like inclusive dihadron production in $e^+ e^-$ annihilations~\cite{Boer:2003ya,Matevosyan:2018icf,Courtoy:2012ry}  and SIDIS. In the following, we outline some interesting applications involving proton-proton collisions and the SIDIS process.

%%%%%%%%%%%%%%
\section{Opportunities with hadron-in-jet fragmentation}
\label{s:chances}

In this section, we mention two possible applications of the above correspondence where results known for dihadron inclusive production can be formally translated into the cross section for hadron-in-jet fragmentation, opening up new channels for investigating the partonic structure of hadrons. 

%%%%%%%%%% Fig. 5  %%%%%%%%%
\begin{figure}[ht]
\begin{center}
\includegraphics[width=0.48\textwidth]{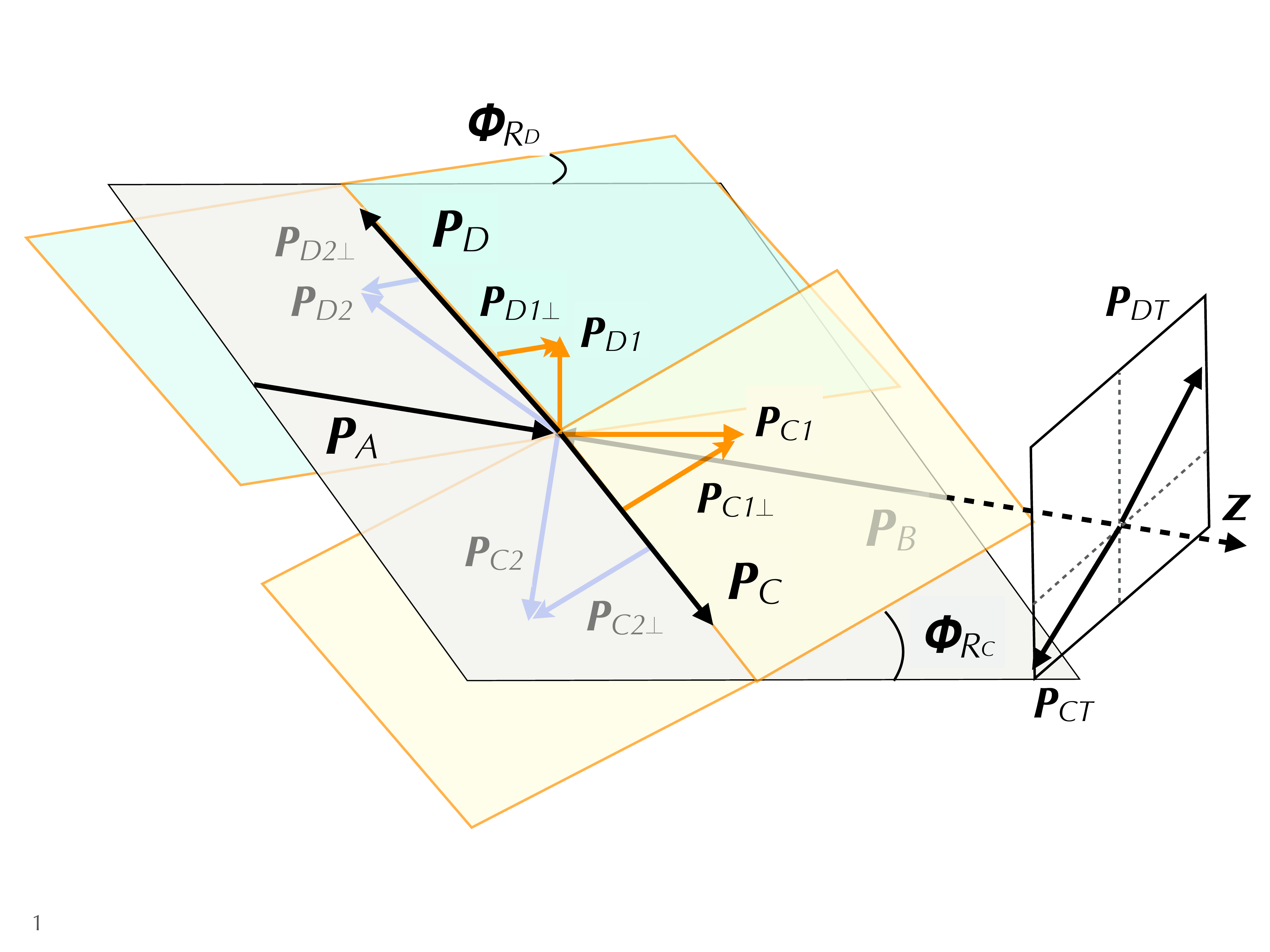} \hspace{0.3cm} 
\includegraphics[width=0.48\textwidth]{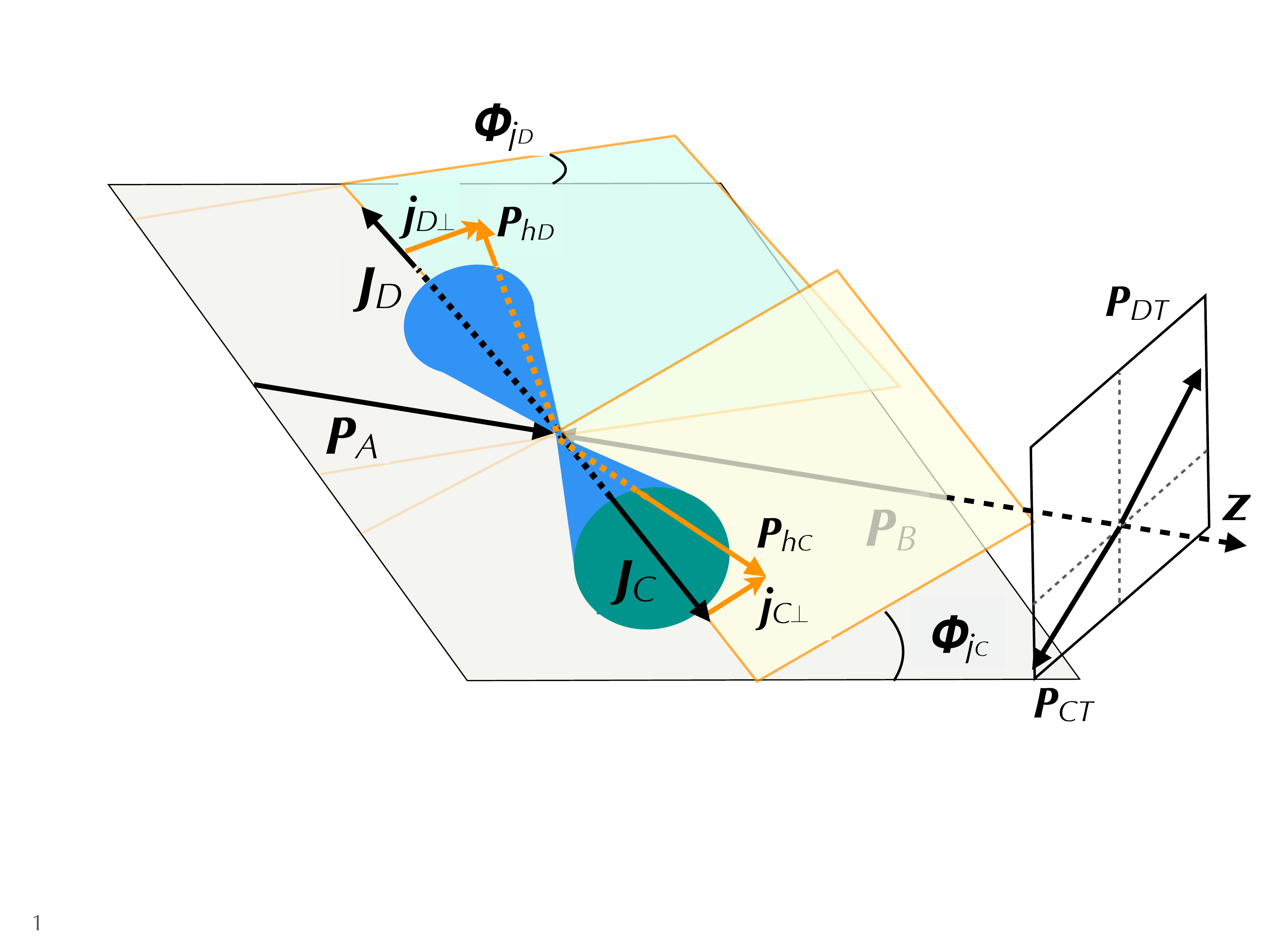} 
\end{center}
\caption{Kinematics for the collision of two unpolarized protons with 3-momenta $\bm{P}_A$ and $\bm{P}_B$ along the $\hat{z}$ axis. Left panel: inclusive production of two back-to-back hadron pairs with total momenta $\bm{P}_C = \bm{P}_{C1} + \bm{P}_{C2}$ and $\bm{P}_D = \bm{P}_{D1} + \bm{P}_{D2}$, and back-to-back projections $\bm{P}_{CT}$ and $\bm{P}_{DT}$ on the transverse plane ($\phi_C = \phi_D + \pi$), respectively; the planes 
containing 
%formed by 
the momenta of each pair 
%are oriented by the 
form the azimuthal angles $\phi_{R_C}$ and $\phi_{R_D}$ with 
%respect to 
the reaction plane 
%formed by
containing
$\bm{P}_A$ and $\bm{P}_C$. Right panel: inclusive production of two back-to-back jets with axis $\hat{\bm{J}}_C$, $\hat{\bm{J}}_D$, and back-to-back projected momenta $\bm{P}_{CT}$ and $\bm{P}_{DT}$ on the transverse plane ($\phi_C = \phi_D + \pi$), respectively; in each jet a hadron is detected with 3-momentum $\bm{P}_{h_C}$ ($\bm{P}_{h_D}$) and transverse component $\bm{j}_{C\perp}$ ($\bm{j}_{D\perp}$) with respect to the jet axis $\hat{\bm{J}}_C$ ($\hat{\bm{J}}_D$); the planes 
%formed by 
containing $\hat{\bm{J}}_C$, $\bm{P}_{h_C}$ and  $\hat{\bm{J}}_D$, $\bm{P}_{h_D}$ 
form 
%are oriented by 
the azimuthal angles $\phi_{j_C}$, $\phi_{j_D}$ with 
%respect to 
the reaction plane 
containing 
%formed by 
$\bm{P}_A$ and $\hat{\bm{J}}_C$.}
\label{f:2x2h_kin}
\end{figure}
%%%%%%%%%%%%%%%%%%%%%%%

%%%%%%%%%%%%%
\subsection{Inclusive production of two dihadrons and back-to-back hadron-in-jet's in proton-proton collisions}
\label{s:pp_2x2h}

For the process $A+B \to (C_1 C_2)_C + (D_1 D_2)_D + X$ depicted in the left panel of Fig.~\ref{f:2x2h_kin}, after 
%integrating 
summing over the polarizations of initial hadrons the leading-twist cross section reads (see App.~\ref{a:cross-check} and Eqs.~(20-22) in Ref.~\cite{Bacchetta:2004it}) 
\begin{align}
&\frac{d\sigma_{UU}}{d\eta_C \, d|\bm{P}_{CT}| \, d\zeta_C \, d\bm{R}_{C\perp} \, d\eta_D \, d|\bm{P}_{DT}| \, d\zeta_D \, d\bm{R}_{D\perp}} = \frac{|\bm{P}_{CT}| |\bm{P}_{DT}|}{8\pi^2} \, \sum_{a,b} \int \frac{dx_A dx_B dz_{hhC} dz_{hhD}}{z_{hhC}^2 z_{hhD}^2} \, f_1^a (x_A)\, f_1^b (x_B) \nonumber \\[0.2cm] 
&\quad \times \hat{s} \delta (\hat{s} + \hat{t} + \hat{u}) \, \delta \left( \frac{|\bm{P}_{CT}|}{z_{hhC}} - \frac{|\bm{P}_{DT}|}{z_{hhD}} \right)  \, \delta \left( x_A P_{Az} - x_B P_{Bz} - \frac{P_{Cz}}{z_{hhC}} - \frac{P_{Dz}}{z_{hhD}} \right)  \nonumber \\
&\quad \times \Bigg\{ \sum_{c,d} \Bigg[ \, \frac{d\hat{\sigma}_{ab\to cd}}{d\hat{t}} \, D_1^c (z_{hhC}, \zeta_C, \bm{R}_{C\perp}^2) \, D_1^d (z_{hhD}, \zeta_D, \bm{R}_{D\perp}^2) \nonumber \\
&\qquad \quad + \cos (\phi_{R_C} - \phi_{R_D}) \, \frac{d\Delta\hat{\sigma}_{ab \to c^\uparrow d^\uparrow}}{d\hat{t}} \, \frac{|\bm{R}_{C\perp}|}{M_{C}} \, H_1^{\sphericalangle \, c} (z_{hhC}, \zeta_C, \bm{R}_{C\perp}^2) \, \frac{|\bm{R}_{D\perp}|}{M_{D}} \, H_1^{\sphericalangle \, d} (z_{hhD}, \zeta_D, \bm{R}_{D\perp}^2) \, \Bigg] \nonumber \\
& \qquad + \cos (2\phi_{R_C} - 2\phi_{R_D}) \,  \frac{d\Delta\hat{\sigma}_{ab \to g^\uparrow g^\uparrow}}{d\hat{t}} \,  
\frac{|\bm{R}_{C\perp}|^2}{M_{C}^2} \,  H_1^{\sphericalangle \, g} (z_{hhC}, \zeta_C, \bm{R}_{C\perp}^2) \,  \frac{|\bm{R}_{D\perp}|^2}{M_{D}^2} \, H_1^{\sphericalangle \, g} (z_{hhD}, \zeta_D, \bm{R}_{D\perp}^2) \Bigg\}  \; , 
\label{e:pp2x2h_xsec}
\end{align}
where the momenta and the angles of the second hadron pair are defined in complete analogy with the first pair by replacing the labels $c,C$ with $d,D$. The delta functions describe in the elementary process the conservation of energy and of momentum both along the longitudinal direction of the $\hat{z}$ axis (identified with $\bm{P}_A$, see Fig.~\ref{f:2x2h_kin}) and in the transverse plane. 

In Eq.~\eqref{e:pp2x2h_xsec}, the elementary cross sections $d\Delta\hat{\sigma}_{ab \to c^\uparrow d^\uparrow}$ involve only quarks for the final partons $c, d$, while $d\Delta\hat{\sigma}_{ab \to g^\uparrow g^\uparrow}$ contain only final gluons linearly polarized in the transverse plane. Hence, the $H_1^{\sphericalangle g}$ function describes the fragmentation of such linearly polarized gluons into pairs of unpolarized hadrons.\footnote{The $H_1^{\sphericalangle \, g}$ corresponds to the notation $\delta \hat{G}^{\sphericalangle}$ of Ref.~\cite{Bacchetta:2004it}.}  %\textcolor{red}{Ale: I wonder if we should reconsider naming this function in a more modern way, i.e, $H_1^{\sphericalangle g}$} \cLR{I decide to use $H_1^{\sphericalangle g}$ by adding a footnote for clarity. Moreover I change the notation everywhere in the article.} 
For both cases of final polarized quarks and gluons, all nonvanishing combinations are listed in the Appendix of Ref.~\cite{Bacchetta:2004it}. 

Therefore, by disentangling specific asymmetries in the azimuthal orientation of the planes containing the momenta of the two dihadrons one can access the DiFFs $H_1^{\sphericalangle \, q}$ and $H_1^{\sphericalangle \, g}$ for the  fragmentation of transversely polarized quarks and linearly polarized gluons, respectively, without considering any polarization in the initial hadronic collision~\cite{Bacchetta:2004it}. Because of the correspondence described in Sec.~\ref{s:pp2h_h-in-jet}, it is interesting to explore the same possibility for the process $A + B \to ({\rm Jet} \, C) + ({\rm Jet} \, D) + X$, as depicted in the right panel of Fig.~\ref{f:2x2h_kin}. Using the same rules of correspondence as in the single hadron-pair production, we get
\begin{align}
&\frac{d\sigma_{UU}}{d\eta_C \, d|\bm{P}_{CT}| \, dz_{hC} \, d\bm{j}_{C\perp} \, d\eta_D \, d|\bm{P}_{DT}| \, dz_{hD} \, d\bm{j}_{D\perp}} = 2 |\bm{P}_{CT}| |\bm{P}_{DT}| \, \sum_{a,b} \int \frac{dx_A dx_B dz_{JC} dz_{JD}}{z_{JC}^2 z_{JD}^2} \, f_1^a (x_A)\, f_1^b (x_B) \nonumber \\[0.2cm] 
&\quad \times \hat{s} \delta (\hat{s} + \hat{t} + \hat{u}) \, \delta \left( \frac{|\bm{P}_{CT}|}{z_{JC}} - \frac{|\bm{P}_{DT}|}{z_{JD}} \right)  \, \delta \left( x_A P_{Az} - x_B P_{Bz} - \frac{P_{Cz}}{z_{JC}} - \frac{P_{Dz}}{z_{JD}} \right)  \nonumber \\
&\quad \times \Bigg\{ \sum_{c,d} \Bigg[ \, \frac{d\hat{\sigma}_{ab\to cd}}{d\hat{t}} \, {\cal D}_1^c (z_{JC}, z_{hC}, \bm{j}_{C\perp}^2; |\bm{P}_{CT}|, |\bm{P}_{CT}| r_C) \, {\cal D}_1^d (z_{JD}, z_{hD}, \bm{j}_{D\perp}^2; |\bm{P}_{DT}|, |\bm{P}_{DT}| r_D) \nonumber \\
&\hspace{2cm} + \cos (\phi_{j_C} - \phi_{j_D}) \, \frac{d\Delta\hat{\sigma}_{ab \to c^\uparrow d^\uparrow}}{d\hat{t}} \, \frac{|\bm{j}_{C\perp}|}{M_{h_C}} \, {\cal H}_1^{\perp \, c} (z_{JC}, z_{hC}, \bm{j}_{C\perp}^2; |\bm{P}_{CT}|, |\bm{P}_{CT}| r_C) \nonumber \\
&\hspace{6.4cm} \times \frac{|\bm{j}_{D\perp}|}{M_{h_D}} \, {\cal H}_1^{\perp \, d} (z_{JD}, z_{hD}, \bm{j}_{D\perp}^2; |\bm{P}_{DT}|, |\bm{P}_{DT}| r_D) \, \Bigg] \nonumber \\
& \qquad \quad + \cos (2\phi_{j_C} - 2\phi_{j_D}) \,
\frac{d\Delta\hat{\sigma}_{ab \to g^\uparrow g^\uparrow}}{d\hat{t}} \, \frac{|\bm{j}_{C\perp}|^2}{M_{h_C}^2} \, {\cal H}_1^{\perp \, g}  (z_{JC}, z_{hC}, \bm{j}_{C\perp}^2; |\bm{P}_{CT}|, |\bm{P}_{CT}| r_C) \nonumber \\
&\hspace{5.8 cm} \times  \frac{|\bm{j}_{D\perp}|^2}{M_{h_D}^2} \, {\cal H}_1^{\perp \, g}  (z_{JD}, z_{hD}, \bm{j}_{D\perp}^2; |\bm{P}_{DT}|, |\bm{P}_{DT}| r_D) \Bigg\}  \; , 
\label{e:2xh-in-jet_cmp}
\end{align}
where $\phi_{j_C}$ and $\phi_{j_D}$ are the azimuthal angles with respect to the reaction plane of the transverse momenta $\bm{j}_{C\perp}, \, \bm{j}_{D\perp}$ of hadrons inside the jets with radius $r_C$ and $r_D$, respectively. All other variables are defined in complete analogy with the single hadron-in-jet case, identifying each corresponding jet by using the labels $c, C$ and $d, D$. 

From Eq.~\eqref{e:2xh-in-jet_cmp}, we deduce that the $\cos (\phi_{j_C} - \phi_{j_D})$ asymmetry in the azimuthal distribution of the two hadrons inside the two back-to-back jets is generated by two ``Collins-in-jet" effects, one per each jet. This asymmetry allows to isolate back-to-back jets produced by the fragmentation of back-to-back transversely polarized quarks, giving an alternative option to access the Collins function of each hadron inside the corresponding jet. 
Similarly and even more interestingly, extracting the $\cos (2 \phi_{j_C} - 2 \phi_{j_D})$ Fourier component in the azimuthal distribution allows to isolate back-to-back jets produced by the fragmentation of back-to-back linearly polarized gluons, giving access to a new class of fragmentation functions: the ${\cal H}_1^{\perp \,g}$ describe the inclusive production of hadrons inside jets by the fragmentation of linearly polarized gluons, where the hadron transverse momentum $\bm{j}_\perp$ with respect to the jet axis becomes the spin analyzer of the gluon linear polarization in the transverse plane.

%%%%%%%%%% Fig. 5  %%%%%%%%%
\begin{figure}[ht]
\begin{center}
\includegraphics[width=0.4\textwidth]{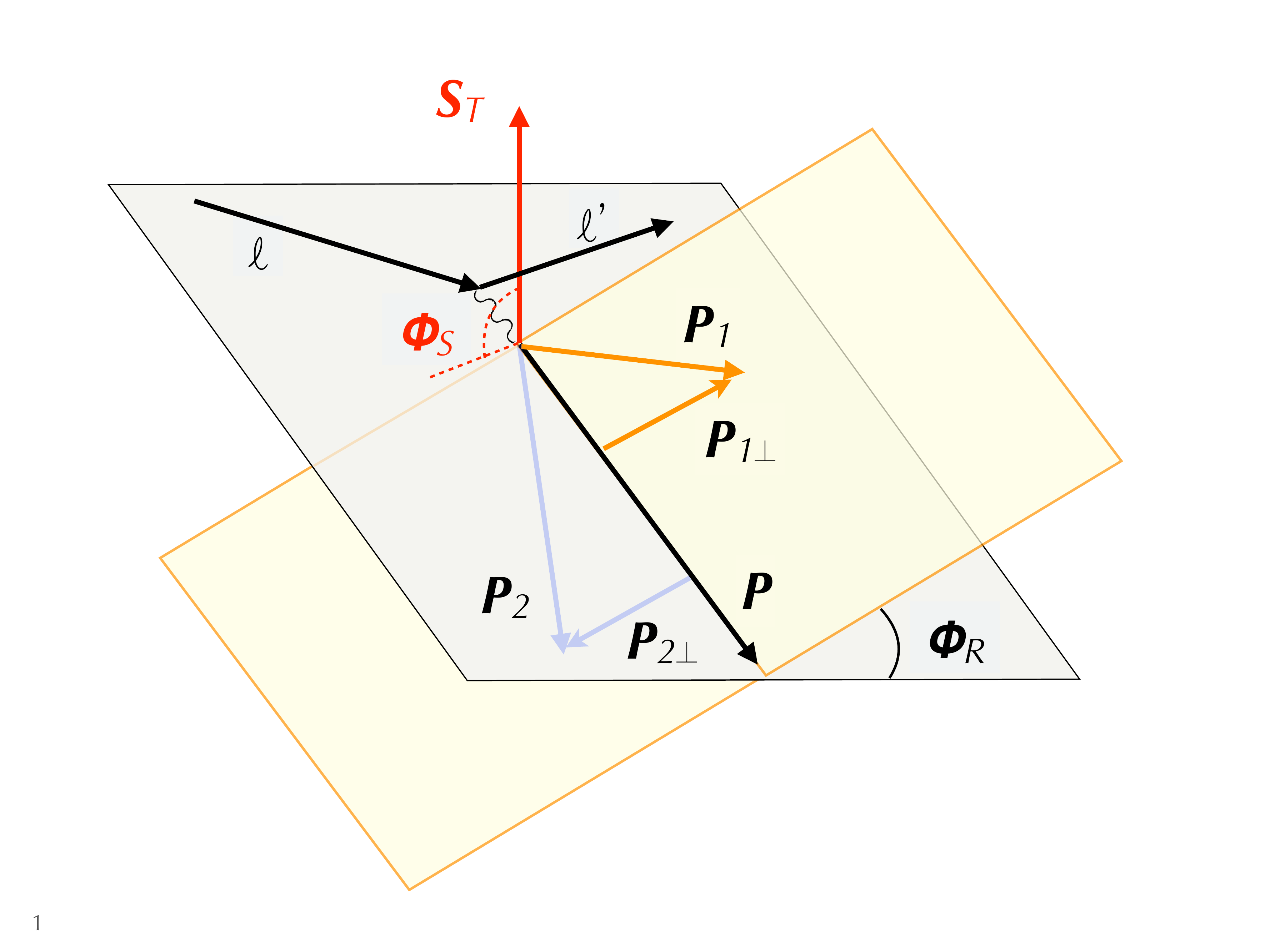} \hspace{0.3cm} 
\includegraphics[width=0.4\textwidth]{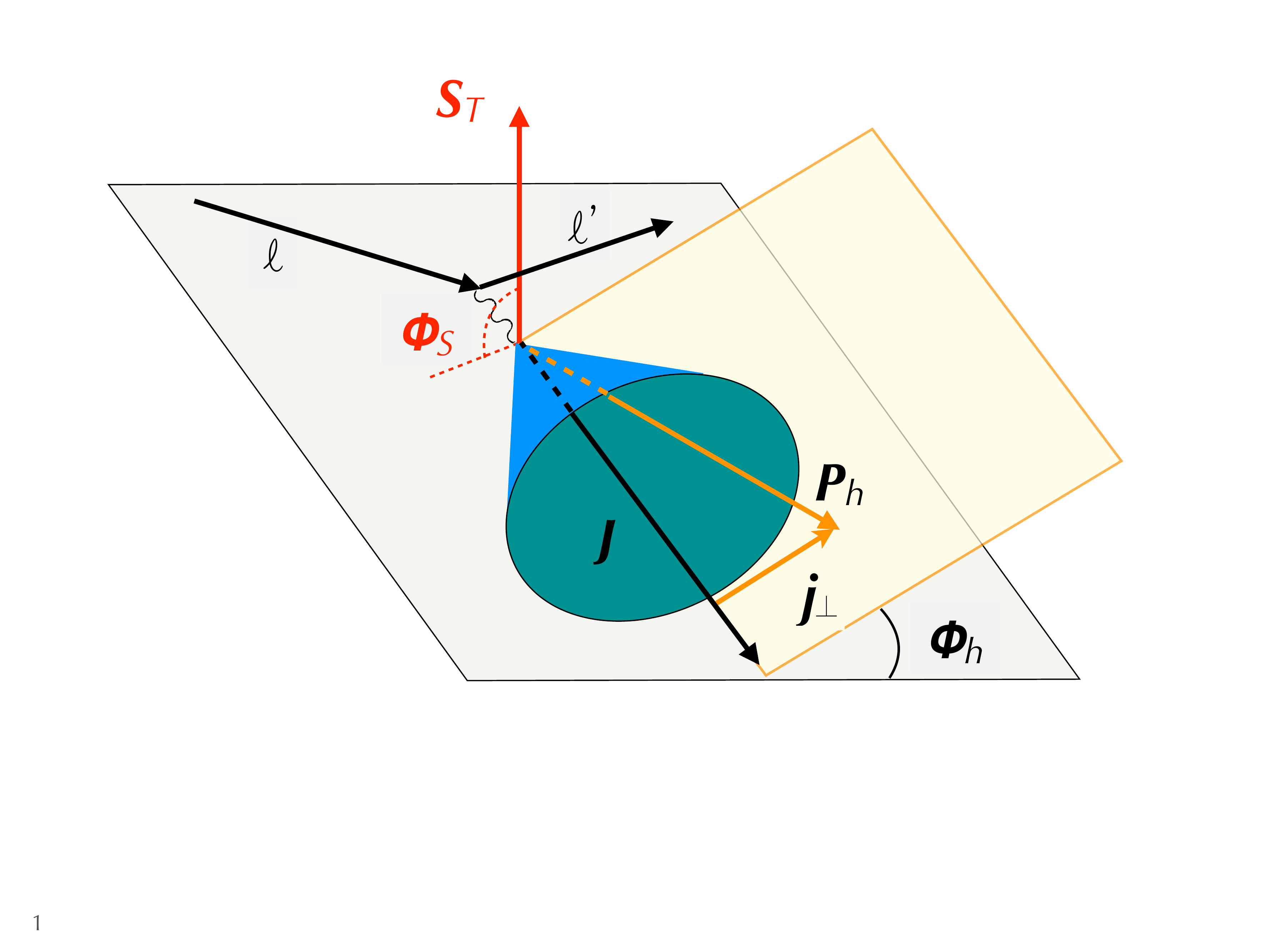} 
\end{center}
\caption{Kinematics for the semi-inclusive deep-inelastic scattering of a lepton with initial momentum $\ell$ and final momentum $\ell'$ on a hadronic  target with transverse polarization $\bm{S}_T$ oriented along $\phi_S$ with respect to the scattering plane formed by $\bm{\ell}$ and $\bm{\ell}'$. The final state can be either a pair of unpolarized hadrons with momenta $P_1$ and $P_2$ with azimuthal orientation $\phi_R$ and total momentum $P = P_1 + P_2$ (left panel), or a hadron with momentum $P_h$ and azimuthal orientation $\phi_h$ inside a jet with standard axis $\hat{\bm{J}}$ (right panel). In both cases, the parallel kinematics is considered where $\bm{P}$ and $\hat{\bm{J}}$ are along the momentum transfer $\bm{q} = \bm{\ell} - \bm{\ell}'$, which identifies the $\hat{z}$ axis.}
\label{f:sidis_kin}
\end{figure}
%%%%%%%%%%%%%%%%%%%%%%%

%%%%%%%%%%%%%
\subsection{Semi-inclusive deep-inelastic scattering up to subleading twist}
\label{s:sidis_twist3}

%A similar correspondence can be established when the pair of hadrons or the hadron inside the jet are produced by the scattering of a lepton with 4-momentum $k$ off a hadronic target with transverse polarization $\bm{S}_T$, leading to a final state with 4-momentum $k'$. The azimuthal orientations of $\bm{S}_T$, of the hadron pair plane (represented by $\bm{R}_\perp = (\bm{P}_{1\perp} - \bm{P}_{2\perp})/2$), and of the hadron inside the jet (represented by $\bm{j}_\perp$), are given by $\phi_S$, $\phi_R$ and $\phi_h$, respectively, which are all measured with respect to the scattering plane identified by $\bm{k}$ and $\bm{k}'$ (see Fig.~\ref{f:sidis_kin}). The hard scale of the scattering is given by $Q^2 = - q^2 = - (k - k')^2 \geq 0$. The collinear kinematics is realized by integrating over the transverse components of the hadron pair total 3-momentum $\bm{P}$, or over the transverse components of the standard jet axis 3-vector $\hat{\bm{J}}$. In both cases, this is equivalent to take $\bm{P}$ or $\hat{\bm{J}}$ collinear with $\hat{\bm{q}}$, which is identified with the $\hat{z}$ axis.  

%At leading twist, the cross section for the inclusive production of a pair of unpolarized hadrons (left panel of Fig.~\ref{f:sidis_kin}) reads~\cite{Bacchetta:2012ty,Radici:2015mwa}

The left panel of Fig.~\ref{f:sidis_kin} describes the kinematics for the $\ell + A \to \ell' + (C_1 C_2) + X$ process, namely for the inclusive production of a hadron pair with momenta $P_1$ and $P_2$ by the scattering of a lepton with 4-momentum $\ell$ off a hadronic target with momentum $P_A$, mass $M$ and polarization $S$, leading to a final lepton with 4-momentum $\ell'$. The azimuthal orientations of the transverse polarization $\bm{S}_T$ and of the hadron pair plane (represented by $\bm{R}_\perp = (\bm{P}_{1\perp} - \bm{P}_{2\perp})/2$) are given by $\phi_S$ and $\phi_R$, respectively, and they are all measured with respect to the scattering plane identified by $\bm{\ell}$ and $\bm{\ell}'$. The hard scale of the process is given by $Q^2 = - q^2 = - (\ell - \ell')^2 \gg M \geq 0$. The collinear kinematics is realized by integrating over the transverse components of the hadron-pair total 3-momentum $\bm{P} = \bm{P}_1 + \bm{P}_2$ or, equivalently, by taking $\bm{P}$ collinear with $\hat{\bm{q}}$, which identifies the $\hat{z}$ axis. 

The expressions of the LO cross section for various combinations of polarization of lepton probe and proton target are listed in Eqs.~(44-49) of Ref.~\cite{Bacchetta:2003vn} up to subleading twist. A slightly different notation was employed in Ref.~\cite{Gliske:2014wba}, where the cross section was described in terms of structure functions, 
%as discussed in ~\cite{Gliske:2014wba}, 
based on the analgous expression in Ref.~\cite{Bacchetta:2006tn}. Here, we limit ourselves to reproducing the terms that are more interesting for our discussion:
%\everymath{\color{red}}
\begin{equation}
\begin{split}
\frac{d\sigma}{dx \, dy \,dz_{hh} \, d\phi_S \, d \bm{R}_{\perp} d\zeta} =
\frac{2 \alpha^2}{x y Q^2}\,
\frac{y^2}{2\,(1-\varepsilon)}\,
%\nonumber \\[0.2em]
% & \quad \times 
 & \biggl\{ F_{UU ,T} + \ldots + |\bm{S}_T|\; \bigg[ 
\varepsilon\, \sin(\phi_R+\phi_S)\, F_{UT}^{\sin\left(\phi_R +\phi_S\right)} + \ldots \bigg] \\
 & \quad + S_{L} \bigg[ \sqrt{2\, \varepsilon (1+\varepsilon)}\, \sin\phi_R\, F_{UL}^{\sin\phi_R} +  \ldots \bigg] 
 \\
%+ 
%\varepsilon
%F_{UU ,L}
%+ \sqrt{2\,\varepsilon (1+\varepsilon)}\,\cos\phi_R\,
%F_{UU}^{\cos\phi_R}
%+ \varepsilon \cos(2\phi_R)\, 
%F_{UU}^{\cos 2\phi_R}
%\nonumber \\
%  & \quad \qquad
& \quad + \lambda\, \sqrt{2\,\varepsilon (1-\varepsilon)}\, \sin\phi_R\, F_{LU}^{\sin\phi_R} + \ldots 
%\phantom{\bigg[ \bigg] } \\
 %& \quad + S_{L}, \bigg[ \sqrt{2\, \varepsilon (1+\varepsilon)}\, \sin\phi_R\, F_{UL}^{\sin\phi_R} +  \ldots
%\varepsilon \sin(2\phi_R)\, 
%F_{UL}^{\sin 2\phi_R}
%\bigg]
%\nonumber \\
% & \quad \qquad
%+ S_\parallel \lambda_e\, \bigg[ \,
%  \sqrt{1-\varepsilon^2}\; 
%F_{LL}
%+\sqrt{2\,\varepsilon (1-\varepsilon)}\,
%  \cos\phi_R\, 
%F_{LL}^{\cos \phi_R}
%\bigg]
%\\
 %& \quad + |\bm{S}_T|\; \bigg[
% \sin(\phi_R-\phi_S)\,
%\Bigl(F_{UT ,T}^{\sin\left(\phi_R -\phi_S\right)}
%+ \varepsilon\, F_{UT ,L}^{\sin\left(\phi_R -\phi_S\right)}\Bigr)
%\nonumber \\ 
% & \qquad  \qquad \qquad
%+ 
%\varepsilon\, \sin(\phi_R+\phi_S)\, F_{UT}^{\sin\left(\phi_R +\phi_S\right)} + \ldots \bigg]
%+ \varepsilon\, \sin(3\phi_R-\phi_S)\,
%F_{UT}^{\sin\left(3\phi_R -\phi_S\right)}
%\phantom{\bigg[ \bigg] }
%\nonumber \\ 
% & \qquad \qquad \qquad
%+ \sqrt{2\,\varepsilon (1+\varepsilon)}\, 
%  \sin\phi_S\, 
%F_{UT}^{\sin \phi_S }
%+ \sqrt{2\,\varepsilon (1+\varepsilon)}\, 
%  \sin(2\phi_R-\phi_S)\,  
%F_{UT}^{\sin\left(2\phi_R -\phi_S\right)}
%\bigg]
%\nonumber \\ 
% & \quad \qquad 
+ |\bm{S}_T|\, \lambda\, \bigg[ \ldots
%  \sqrt{1-\varepsilon^2}\, \cos(\phi_R-\phi_S)\, 
%F_{LT}^{\cos(\phi_R -\phi_S)}
%+\sqrt{2\,\varepsilon (1-\varepsilon)}\, 
%  \cos\phi_S\, 
%F_{LT}^{\cos \phi_S}
%\nonumber \\ 
% & \qquad \qquad \qquad
%+\sqrt{2\,\varepsilon (1-\varepsilon)}\, 
%  \cos(2\phi_R-\phi_S)\,  
%F_{LT}^{\cos(2\phi_R - \phi_S)}
\bigg] \biggr\} \,,
\label{e:crossmaster}
\end{split}
\end{equation}
%\everymath{\color{black}}
%
%Here, we focus on the combination of longitudinally polarized probe and unpolarized target ($d\sigma_{LU}$ of Eq.~(47) in Ref.~\cite{Bacchetta:2003vn}), and of unpolarized probe and longitudinally polarized target ($d\sigma_{UL}$ of Eq.~(45) in Ref.~\cite{Bacchetta:2003vn}). For convenience, we list the cross sections below: 
%\begin{align}
% \frac{d\sigma_{LU}}{d\zeta d\bm{R}_\perp dx dz_{hh} dy d\phi_S} &= \frac{\alpha^2}{2 \pi x y Q^2} \lambda \sum_q x e_q^2 \, \frac{W(y)}{2}  \sin \phi_R \frac{2 |\bm{R}_\perp|}{Q} \Bigg[ \frac{M}{M_{hh}} x e^q(x) \, H_1^{\sphericalangle \, q} (z_{hh},\zeta, \bm{R}_\perp^2) + \frac{1}{z_{hh}} f_1^q (x) \, \tilde{G}^{\sphericalangle \, q} (z_{hh},\zeta, \bm{R}_\perp^2) \Bigg] \; ,   \label{e:SIDISdsigLU} 
 %\\[0.5cm]
%\frac{d\sigma_{UL}}{d\zeta d\bm{R}_\perp dx dz_{hh} dy d \phi_S} &= \frac{\alpha^2}{2 \pi x y Q^2} S_L \sum_q x e_q^2 \, \frac{V(y)}{2} \sin \phi_R \frac{2 |\bm{R}_\perp|}{Q} \Bigg[ \frac{M}{M_{hh}} x h_L^q (x) \, H_1^{\sphericalangle \, q} (z_{hh},\zeta, \bm{R}_\perp^2) + \frac{1}{z_{hh}} g_1^q(x) \, \tilde{G}^{\sphericalangle \, q} (z_{hh},\zeta, \bm{R}_\perp^2) \Bigg] \; ,  \label{e:SIDISdsigUL} 
%\end{align}
where $\alpha$ is the fine structure constant, $x = Q^2 / 2 P_A \cdot q \approx k^+ / P_A^+$ is the fraction of target momentum carried by a parton with 4-momentum $k$ and fractional charge $e_q$, $y = P_A \cdot q / P_A \cdot \ell \approx (E_{\ell} - E'_{\ell}) / E_{\ell}$ is the fraction of beam energy transferred to the hadronic system, $\lambda$ is the beam helicity, $S_L$ is the target longitudinal polarization, $\varepsilon$ is the ratio of longitudinal and transverse photon flux, and 
%\everymath{\color{red}}
\begin{align} 
\frac{y^2}{2\,(1-\varepsilon)} &\approx \left(1-y +\frac{1}{2} y^2\right),
&
\frac{y^2}{2\,(1-\varepsilon)}\,\varepsilon &\approx (1-y)
\\
\frac{y^2}{2\,(1-\varepsilon)}\,\sqrt{2\,\varepsilon(1+\varepsilon)} &\approx (2-y)\, \sqrt{1-y},
&
\frac{y^2}{2\,(1-\varepsilon)}\,\sqrt{2\,\varepsilon(1-\varepsilon)} &\approx y \,  \sqrt{1-y} \; .
\end{align}
%\begin{equation}
%\frac{W(y)}{2} = y \sqrt{1-y} \; , \qquad %\frac{V(y)}{2} = (2 - y) \sqrt{1-y} \; .
%\label{e:SIDISdepol}
%\end{equation}
%\textcolor{red}{
%Ale: instead of writing the full cross section, I would prefer to write the structure functions, as defined in Eq.(29) of our paper with Gliske. However, prefactors need to be checked. First of all, I don't know why there's a factor $W(x,y)/2$ instead of just $W(x,y)$  in Gliske. Secondly, in Gliske we quote the cross section differential in $dM_h$ instead of $dM_h^2$, so there should be a $2 M_h$ factor difference, which however I don't see in, e.g., Eq.~(44) of Gliske for $l=1$, $m=1$ }
%\everymath{\color{red}}

The structure functions of interest can be written in terms of PDFs and DiFFs in the following way
\begin{align}
%\begin{split}
 \label{e:SIDISFUT}
F_{UT}^{\sin (\phi_R + \phi_S)} &=  \frac{1}{4 \pi} \, x \sum_q  \, e_q^2 \, \frac{|\bm{R}_\perp|}{M_{hh}} \, h_1^q(x) \, {H}_1^{\sphericalangle \, q} (z_{hh},\zeta, \bm{R}_\perp^2) \; , 
\\
%\end{split}
%\begin{split}
\label{e:SIDISFLU}
 F_{LU}^{\sin \phi_R} &= \frac{2 M}{Q} \, \frac{1}{4 \pi} \, x \sum_q \, e_q^2 \, 
\frac{|\bm{R}_\perp|}{M_{hh}}\Bigg[ e^q(x) \, H_1^{\sphericalangle \, q} (z_{hh},\zeta, \bm{R}_\perp^2) + \frac{M_{hh}}{M} \, f_1^q (x) \, \frac{\tilde{G}^{\sphericalangle \, q} (z_{hh},\zeta, \bm{R}_\perp^2)}{z_{hh}} \Bigg], 
%\end{split}
\\
%\begin{split}
\label{e:SIDISFUL}
F_{UL}^{\sin \phi_R} &= \frac{2 M}{Q} \, \frac{1}{4 \pi} \, x \sum_q \, e_q^2 \, 
\frac{|\bm{R}_\perp|}{M_{hh}} \Bigg[ h_L^q (x) \, H_1^{\sphericalangle \, q} (z_{hh},\zeta, \bm{R}_\perp^2) + \frac{M_{hh}}{M} \, g_1^q(x) \,\frac{ \tilde{G}^{\sphericalangle \, q} (z_{hh},\zeta, \bm{R}_\perp^2)}{z_{hh}} \Bigg] \; .
%\end{split}
%\\
\end{align}
%\everymath{\color{black}}
%
%\textcolor{red}{Ale: I agree to emphasize the $e(x)$ access, but we should also mention transversity and include also}
%\everymath{\color{red}}
%\begin{equation}
%\begin{split}
%\frac{d\sigma_{UT}}{d\zeta d\bm{R}_\perp dx dz_{hh} dy d\phi_S} & = \frac{\alpha^2}{ 2 \pi x y Q^2} \, \vert S_{hT}\vert  \sum_q \, x e_q^2 \, \Bigg\{ B(y) \, \sin (\phi_R + \phi_S) \, \frac{|\bm{R}_\perp|}{M_h} h_1^q(x) \, { H}_1^{\sphericalangle \, q} 
%(z_{hh},\zeta, \bm{R}_\perp^2) \
%\\
%& \quad +\frac{V(y)}{2} \sin \phi_S \, \frac{2 M_h}{Q} \Bigg[h_1^q(x) \, \frac{1}{z_{hh}}
%\tilde{H}^{q} (z_{hh},\zeta, \bm{R}_\perp^2)  
%-\frac{M}{M_h} \, x f_T^q (x) \, {D}_1^{q}
%(z_{hh},\zeta, \bm{R}_\perp^2)  \Bigg]\bigg\}
%\end{split}
%\end{equation}
%\textcolor{red}{However, I am not sure about the $\tilde{H}$ term in the last line. We have an extra term in Eq.~(46) of our 2003 paper, and we have something apparently different in our paper with Gliske, Eq.~(54) for $l=0$ and $m=0$ (we probably forgot the superscript in $\tilde{H}^{| l,m \rangle}$).
%If we use the structure functions, it should easier and can be obtained also from Gliske, Eq.~(52), with $l=1$, $m=1$:}
%\begin{align}

%\end{align}
%\everymath{\color{black}}

Equation~\eqref{e:SIDISFUT} represents the standard way to address in a collinear framework the chiral-odd transversity PDF $h_1(x)$. The integral of $h_1(x)$ is the tensor charge, which might represent a possible portal to new physics beyond the Standard Model~\cite{Courtoy:2015haa} since it is relevant for explorations of new possible CP-violating couplings~\cite{Dubbers:2011ns} or effects induced by tensor operators not included in the Standard Model Lagrangian~\cite{Bhattacharya:2011qm}. The tensor charge can be computed in lattice QCD with very high precision~\cite{Constantinou:2020hdm}. Future facilities will have a large impact on the current uncertainty on the tensor charge extracted from phenomenological studies~\cite{Proceedings:2020fyd,AbdulKhalek:2021gbh,AbdulKhalek:2022hcn}. 

Equation~\eqref{e:SIDISFLU} is particularly interesting because it contains the contribution of the twist-3 chiral-odd PDF  $e(x)$, which contains crucial information on quark-gluon-quark correlations (see, e.g., \cite{Efremov:2002qh}). The integral of $e(x)$ is the scalar charge of the nucleon and is related to the 
%In fact, its isoscalar Mellin moment at $Q^2 = 0$ gives 
the so-called $\sigma$ term, which plays an important role in understanding the emergence of nucleon mass from chiral symmetry breaking~\cite{Jaffe:1991kp} and its decomposition in terms of contributions from quarks and gluons~\cite{Ji:1994av,Ji:2020baz,Lorce:2017xzd,Lorce:2021xku}. The nucleon scalar charge can be important also for the search of physics beyond the Standard Model, since it probes scalar interactions and can be relevant for dark matter searches (see, e.g., Refs.~\cite{Bhattacharya:2011qm,Ellis:2008hf}).  The nucleon scalar charge and the $\sigma$ term have been computed in lattice QCD (for a review, see Ref.~\cite{FlavourLatticeAveragingGroupFLAG:2021npn} and references therein). 
The $e(x)$ has been studied in several non-perturbative models of hadron structure~\cite{Jaffe:1991ra,Wakamatsu:2000fd,Schweitzer:2003uy,Cebulla:2007ej,Mukherjee:2009uy,Avakian:2010br,Lorce:2014hxa,Pasquini:2018oyz,Bastami:2020rxn}. It can be extracted in the TMD framework by considering the beam spin asymmetry that isolates the $d\sigma_{LU}$ cross section for inclusive single-hadron production, where the chiral-odd partner is represented by the Collins function $H_1^\perp$~\cite{Mulders:1995dh,Efremov:2002ut,Avakian:2003pk}.  
%\cLR{I think that the we have to cancel or rewrite the next sentences, because by changing the notation we don't have yet the extra terms expect to the gluon one.}
However, 
%as in Eq.~\eqref{e:SIDISFLU} 
this observable contains other three contributions~\cite{Bacchetta:2004zf,Yuan:2003gu,Gamberg:2003pz}. Moreover, each term is represented by an intricate convolution upon transverse momenta. Therefore, it may be more convenient to work in the collinear framework and isolate the $e(x)$ through the simple product with its chiral-odd partner represented by the DiFF $H_1^{\sphericalangle}$, as shown in Eq.~\eqref{e:SIDISFLU}.

In order to reach this goal, we need to deal with the second contribution in Eq.~\eqref{e:SIDISFLU}, which depends on the unknown twist-3 DiFF $\tilde{G}^{\sphericalangle}$. Calculations in the spectator model show that $\tilde{G}^{\sphericalangle}$ turns out to be small, and possibly with opposite sign to $H_1^{\sphericalangle}$~\cite{Yang:2019aan}. The extraction of $e(x)$ from CLAS and CLAS12 data projected onto the $x$ dependence was performed assuming that the $M_{hh}$ dependence of $\tilde{G}^{\sphericalangle}$ is the same of $H_1^{\sphericalangle}$ but rescaled by a constant factor~\cite{Courtoy:2022kca}. A possible strategy to overcome this problem could be to study the ratio $d\sigma_{LU} / d\sigma_{UL}$~\cite{Pisano:2015wnq}. In fact, if the term proportional to $\tilde{G}^{\sphericalangle}$ would be negligible, using the flavor symmetries of $H_1^{\sphericalangle}$~\cite{Bianconi:1999uc,Bacchetta:2006un,Albino:2008aa,Bacchetta:2011ip,Courtoy:2012ry,Bacchetta:2012ty,Radici:2015mwa,Radici:2016lam,Radici:2018iag} the ratio should not exhibit any dependence on $(z_{hh}, M_{hh})$, since the latter should cancel out between numerator and denominator. On the contrary, any observed dependence would hint at a non-negligible contribution from twist-3 DiFF, making the extraction of $e(x)$ more challenging. 

In this perspective, collecting more information on these observables from other channels should give more insight. Hence, it might be useful to consider the $\ell + A \to \ell' + (\mathrm{Jet} \, C) + X$ process depicted in the right panel of Fig.~\ref{f:sidis_kin}, namely the SIDIS on a (polarized) hadron target where a hadron with momentum $P_h$ is inclusively produced inside a jet with transverse momentum $\bm{j}_\perp$ with respect to the jet axis $\hat{\bm{J}}$ taken parallel to the $\hat{z} = \hat{\bm{q}}$ axis. The cross section of this process has the same structure of Eq.~\eqref{e:crossmaster}. By using the correspondence of Sec.~\ref{s:pp2h_h-in-jet}, the structure functions in Eqs.~\eqref{e:SIDISFUT}-\eqref{e:SIDISFUL} become
%\begin{align}
%\frac{d\sigma_{LU}}{dz_h d\bm{j}_\perp dx dz_J dy d\phi_S} & = \frac{2 \alpha^2}{x y Q^2} \, \lambda \sum_q \, x e_q^2 \, \frac{W(y)}{2} \, \sin \phi_j \, \frac{2|\bm{j}_\perp|}{Q} \Bigg[ \frac{M}{M_h} \, x e^q (x) \, {\cal H}_1^{\perp \, q} (z_J,  z_h, \bm{j}_\perp^2; Q, Q r) \nonumber \\
%&\hspace{5.5cm} + \frac{1}{z_h} \, f_1^q (x) \, \tilde{\cal G}^{\perp \, q} (z_J,  z_h, \bm{j}_\perp^2; Q, Q r) \Bigg] \;, \label{e:JdsigLU} \\[0.2cm]
%\frac{d\sigma_{UL}}{dz_h d\bm{j}_\perp dx dz_J dy d\phi_S} & = \frac{2 \alpha^2}{x y Q^2} \, S_L \sum_q \, x e_q^2 \, \frac{V(y)}{2} \, \sin \phi_j \, \frac{2|\bm{j}_\perp|}{Q} \, \Bigg[ \frac{M}{M_h} \, x h_L^q (x) \, {\cal H}_1^{\perp \, q} (z_J,  z_h, \bm{j}_\perp^2; Q, Q r) \nonumber \\
%&\hspace{5.5cm} + \frac{1}{z_h} \, g_1^q (x) \, \tilde{\cal G}^{\perp \, q} (z_J,  z_h, \bm{j}_\perp^2; Q, Q r) \Bigg] \;, \label{e:JdsigUL} 
%\end{align}
%\textcolor{red}{Ale: same as before: instead of writing the full cross section, I would prefer to write the structure functions:}
%\everymath{\color{red}}
\begin{align}
F_{UT}^{\sin (\phi_j + \phi_S)} &=  x \, \sum_q  \, e_q^2 \, \frac{|\bm{j}_\perp|}{M_h} \, h_1^q(x) \, {\cal H}_1^{\perp \, q} (z_J,  z_h, \bm{j}_\perp^2; Q, Q r) \; , \label{e:jetSIDISFUT}
\\ 
F_{LU}^{\sin \phi_j} &=  \frac{2M}{Q} \, x \, \sum_q \, e_q^2 \, 
\frac{|\bm{j}_\perp|}{M_{h}} \Bigg[ e^q (x) \, {\cal H}_1^{\perp \, q} (z_J,  z_h, \bm{j}_\perp^2; Q, Q r) 
+ \frac{M_{h}}{M} \, f_1^q (x) \, \frac{\tilde{\cal G}^{\perp \, q} (z_J,  z_h, \bm{j}_\perp^2; Q, Q r)}{z_{h}} \Bigg] \; , \label{e:jetSIDISFLU}
\\
F_{UL}^{\sin \phi_j} &= \frac{2M}{Q} \, x \, \sum_q \, e_q^2 \, 
\frac{|\bm{j}_\perp|}{M_h} \, \Bigg[ h_L^q (x) \, {\cal H}_1^{\perp \, q} (z_J,  z_h, \bm{j}_\perp^2; Q, Q r) + \frac{M_h}{M} \, g_1^q (x) \, \frac{\tilde{\cal G}^{\perp \, q} (z_J,  z_h, \bm{j}_\perp^2; Q, Q r)}{z_h} \Bigg] \; , \label{e:jetSIDISFUL}
\end{align}
%\everymath{\color{black}}
%\cLR{I write the four previous equations in a similar way to the di-hadron ones. In particular the structure functions now are in the same form as the ones in the new testament.} \\
where $Q r$ is the typical scale of the jet with radius $r$. Equation~\eqref{e:jetSIDISFUT} indicates a new way to address the collinear transversity PDF $h_1(x)$, namely through the ``Collins-in-jet" effect in the SIDIS process. Equations~\eqref{e:jetSIDISFLU},\eqref{e:jetSIDISFUL} show that in the same framework of the SIDIS ``Collins-in-jet" effect one can address also the twist-3 collinear PDFs $e(x)$ and $h_L(x)$, provided that the remaining contribution given by $\tilde{\cal G}^\perp$ is small. The $\tilde{\cal G}^\perp$ is a new twist-3 jTMDFF that corresponds to the above twist-3 DiFF $\tilde{G}^\sphericalangle$. As in the dihadron case, by using the current knowledge on the ``Collins-in-jet" effect one could predict the dependence of the ratio $d\sigma_{LU} / d\sigma_{UL}$ on the kinematic variables of the final state. The analysis of any possible deviation of data from these predictions would indicate if the contribution of the twist-3 $\tilde{\cal G}^\perp$ would be or would not be negligible.

\section{Conclusions}
\label{s:end}

Transverse-Momentum-Dependent (TMD) factorization gives the possibility of measuring many interesting signals and access many intriguing features of the structure of hadrons. However, one of its shortcomings is that it cannot be applied to hadronic collisions with observed hadronic final states, like, e.g., the process $A+B \to C + D + X$.

Two alternative mechanisms have been proposed to recover part of the versatility of TMDs while preserving the applicability to hadronic processes: the inclusive production of dihadrons, or of a hadron inside jet. 
The inclusive production of dihadrons, namely of two hadrons originating from the fragmentation of the same parton, can be usefully studied in the collinear framework, where transverse momenta of all partons are integrated; it involves universal collinear Dihadron Fragmentation Functions (DiFFs). The inclusive production of a hadron inside jet, namely the inclusive production of a jet with a detected substructure, can be studied in a hybrid factorization approach involving collinear partonic functions in the initial state and TMD hadron-in-jet Fragmentation Functions (jTMDFFs) in the final state. 

In this paper, we have explored similarities between the two formalisms of dihadron and hadron-in-jet production, and we have established a set of correspondence rules between DiFFs and jTMDFFs. We have used this correspondence to transfer to the jTMDFF case some interesting results obtained with DiFFs, in particular for inclusive production of two back-to-back dihadrons in unpolarized proton-proton collisions, and for inclusive production of a dihadron in semi-inclusive deep-inelastic scattering. 

In unpolarized proton-proton collisions with the inclusive production of two back-to-back jets where one hadron is detected inside each jet, the cross section contains specific modulations that can distinguish if the hadron is detected inside a jet generated by a quark or by a gluon. Moreover, one of the two modulations is sensitive to a polarized jTMDFF directly linked to the TMD fragmentation function of a linearly polarized gluon. 

In semi-inclusive deep-inelastic scattering, the cross section for unpolarized lepton probe and transversely polarized proton target offers a new channel to extract the chiral-odd transversity collinear parton distribution function $h_1(x)$, which is connected to the puzzling proton tensor charge. The cross section for longitudinally polarized lepton and unpolarized proton contains a term proportional to the chiral-odd subleading-twist collinear parton distribution function $e(x)$, which is connected to the well known nucleon $\sigma$ term and to the physics of QCD chiral symmetry breaking. 

The above examples illustrate how useful the formal comparison between DiFFs and jTMDFFs can be. The inclusive production of dihadrons has already been measured in hadronic colliders~\cite{Adamczyk:2015hri,STAR:2017wsi,Pokhrel:2021igk}, $e^+ e^-$ colliders~\cite{Vossen:2011fk} and fixed-target experiments~\cite{Airapetian:2008sk,Adolph:2012nw,Adolph:2014fjw,Braun:2015baa}. The inclusive production of hadrons-in-jet has been measured only in hadronic colliders~\cite{STAR:2017akg,STAR:2022hqg}. Both channels will be (abundantly) available at the future Electron-Ion Collider~\cite{AbdulKhalek:2021gbh,AbdulKhalek:2022hcn}. Therefore, we think it is worth to explore the above mentioned possibilities and to push further the analysis of the consequences of the correspondence rules set in this paper.

%%%%%%%%%%%%%%%%%%%%%%%%%%%%%%%%%%%%%%%%%%%%%%%%%%%%%
%\begin{acknowledgments}

%\end{acknowledgments}

%%%%%%%%%%%%%%%%%%%%%

\appendix

\section{Cross section for inclusive dihadron production}
\label{a:cross-check}

In Ref.~\cite{Bacchetta:2004it}, Eq.~(15) shows the unpolarized cross section for the process $A + B \to (C_1 \, C_2) + X$. After integrating over the azimuthal orientation $\phi_{S_B}$ of the polarization of hadron $B$, it reads
\begin{align}
\frac{d\sigma_{UU}}{d\eta \, d|\bm{P}_T| \, d\cos\theta_C \, dM_{hh}^2 \, d\phi_R} &= \nonumber \\
&\hspace{-1.5cm} = \frac{|\bm{P}_T|}{2\pi} \, \sum_{a,b,c,d} \int \frac{dx_A dx_B dz_{hhC}}{z_{hhC}^2} \, f_1^a (x_A)\, f_1^b (x_B) \, z_{hhC} \delta (z_{hhC} - \bar{z}_{hh}) \, \frac{d\hat{\sigma}_{ab\to cd}}{d\hat{t}} \, D_1^c (z_{hhC}, \cos\theta_C, M_{hh}^2) \nonumber \\
&\hspace{-1.5cm} = \frac{|\bm{P}_T|}{2\pi s} \, \sum_{a,b,c,d} \int \frac{dx_A dx_B dz_{hhC}}{x_A x_B z_{hhC}^2} \, f_1^a (x_A)\, f_1^b (x_B) \, \hat{s}^2 \delta (\hat{s} + \hat{t} + \hat{u}) \, \frac{d\hat{\sigma}_{ab\to cd}}{d\hat{t}} \, D_1^c (z_{hhC}, \cos\theta_C, M_{hh}^2)  \; , 
\label{e:pp2h_PRD}
\end{align}
where $\eta, |\bm{P}_T|,  M_{hh}^2$ and $\phi_R$ are defined in Sec.~\ref{s:pp2h}, and $\theta_C$ is the polar angle between $\bm{P}$ and the direction of the back-to-back emission of the two hadrons in their center-of-mass (c.m.) frame (see Fig.~3 of Ref.~\cite{Bacchetta:2002ux}). 

It turns out that $\zeta = a + b \cos \theta_C$, with $a, b$ functions of only the invariant mass $M_{hh}$~\cite{Bacchetta:2002ux}. Therefore, the Jacobian of the transformation is $d\zeta = 2 |\bm{R}| / M_{hh} \, d\cos \theta_C$ with~\cite{Bacchetta:2004it}
\begin{equation}
|\bm{R}| = \frac{1}{2} \sqrt{M_{hh}^2 - 2 (M_1^2 + M_2^2) + (M_1^2 - M_2^2)^2 / M_{hh}^2} \; .
\label{e:modR}
\end{equation}
Using the kinematic relations in Eq.~\eqref{e:2h_invar} and the obvious definition $\bm{R}_\perp = ( |\bm{R}_\perp| \cos \phi_R, \, |\bm{R}_\perp| \sin \phi_R)$, we can compute the Jacobian of the transformation $dM_{hh}^2 \, d\phi_R = d\bm{R}_\perp \, 8/(1-\zeta^2)$. The cross section in Eq.~\eqref{e:pp2h_PRD} can be conveniently rewritten as
\begin{align}
\frac{d\sigma_{UU}}{d\eta \, d|\bm{P}_T| \, d\zeta \, d\bm{R}_\perp} &=  \sum_{a,b,c,d} \int \frac{dx_A dx_B dz_{hhC}}{x_A x_B z_{hhC}^2} \, f_1^a (x_A)\, f_1^b (x_B) \, 
\frac{|\bm{P}_T|\, \hat{s}}{2\pi s} \, \frac{d\hat{\sigma}_{ab\to cd}}{d\hat{t}} \, \hat{s} \delta(\hat{s} + \hat{t} + \hat{u})  \, D_1^c (z_{hhC}, \zeta, \bm{R}_\perp^2)  \; , 
\label{e:pp2h_cmp}
\end{align}
where 
\begin{equation}
D_1^c (z_{hhC}, \cos\theta_C, M_{hh}^2) = 2 \frac{|\bm{R}|}{M_{hh}} \frac{1-\zeta^2}{8}\,  D_1^c (z_{hhC}, \zeta, \bm{R}_\perp^2)  
\end{equation}
takes into account the above Jacobians. 

In a similar way, Eq.~(16) of Ref.~\cite{Bacchetta:2004it} describes the polarized cross section for the process $A + B^\uparrow \to (C_1 \, C_2) + X$: 
\begin{align}
&\frac{d\sigma_{UT}}{d\eta \, d|\bm{P}_T| \, d\cos\theta_C \, dM_{hh}^2 \, d\phi_R d\phi_{S_B}} = \frac{|\bm{P}_T|}{4\pi^2} \, |\bm{S}_{BT}| \,  \sin (\phi_{S_B} - \phi_R) \nonumber \\ 
&\qquad \times  \sum_{a,b,c,d} \int \frac{dx_A dx_B dz_{hhC}}{z_{hhC}^2} \, f_1^a (x_A)\, h_1^b (x_B) \, z_{hhC} \delta (z_{hhC} - \bar{z}_{hh}) \, \frac{d\Delta\hat{\sigma}_{ab^\uparrow \to c^\uparrow d}}{d\hat{t}} \, \frac{|\bm{R}|}{M_{hh}} \, \sin \theta_C \, H_1^{\sphericalangle \, c} (z_{hhC}, \cos\theta_C, M_{hh}^2) \nonumber \\
&\quad = \frac{|\bm{P}_T|}{4\pi^2 s} \, |\bm{S}_{BT}| \,  \sin (\phi_{S_B} - \phi_R) \nonumber \\
&\qquad \times \sum_{a,b,c,d} \int \frac{dx_A dx_B dz_{hhC}}{x_A x_B z_{hhC}^2} \, f_1^a (x_A)\, h_1^b (x_B) \, \hat{s}^2 \delta (\hat{s} + \hat{t} + \hat{u}) \, \frac{d\Delta\hat{\sigma}_{ab^\uparrow \to c^\uparrow d}}{d\hat{t}} \, \frac{|\bm{R_\perp}|}{M_{hh}} \, H_1^{\sphericalangle c} (z_{hhC}, \cos\theta_C, M_{hh}^2)  \; ,
\label{e:ppT2h_PRD}
\end{align}
where $\bm{S}_{BT}$ is the transverse polarization of the colliding proton with orientation $\phi_{S_B}$ with respect to the reaction plane, and $h_1^b$ is the transversity distribution for the transversely polarized parton $b$ with fractional momentum $x_B$. The elementary cross sections $d\Delta\hat{\sigma}_{ab^\uparrow \to c^\uparrow d}$ describe the annihilation of parton $a$ and $b$ with transfer of the transverse polarization of the latter to parton $c$ while summing on the undetected fragments from parton $d$. All the possible independent flavor combinations are listed in the Appendix of Ref.~\cite{Bacchetta:2004it}. 

By applying the same transformation of variables from $d\cos\theta_C dM_{hh}^2 d\phi_R$ to $d\zeta d\bm{R}_\perp$, we get
\begin{align}
\frac{d\sigma_{UT}}{d\eta \, d|\bm{P}_T| \, d\zeta \, d\bm{R}_\perp d\phi_{S_B}} &= \frac{ |\bm{S}_{BT}| }{4\pi^2}\,  \sin (\phi_{S_B} - \phi_R) \nonumber \\ 
&\hspace{-2cm} \times \sum_{a,b,c,d} \int \frac{dx_A dx_B dz_{hhC}}{x_A x_B z_{hhC}^2} \, f_1^a (x_A)\, h_1^b (x_B) \, 
\frac{|\bm{P}_T|\, \hat{s}}{s} \, \frac{d\Delta \hat{\sigma}_{ab^\uparrow \to c^\uparrow d}}{d\hat{t}} \, \hat{s} \delta(\hat{s} + \hat{t} + \hat{u}) \, \frac{|\bm{R_\perp}|}{M_{hh}} \, H_1^{\sphericalangle c} (z_{hhC}, \zeta, \bm{R}_\perp^2)  \; , 
\label{e:ppT2h_cmp}
\end{align}
where 
\begin{equation}
H_1^{\sphericalangle c} (z_{hhC}, \cos\theta_C, M_{hh}^2) = 2 \frac{|\bm{R}|}{M_{hh}} \frac{1-\zeta^2}{8}\,  H_1^{\sphericalangle c} (z_{hhC}, \zeta, \bm{R}_\perp^2)  \; .
\end{equation}

We generalize the above formulae to the case of the inclusive production of two dihadrons. In Ref.~\cite{Bacchetta:2004it}, from Eqs.~(20-22) the unpolarized cross section for the process $A+B \to (C_1 C_2)_C + (D_1 D_2)_D + X$ reads (after integrating on the polarizations of initial hadrons)
\begin{align}
&\frac{d\sigma_{UU}}{d\eta_C \, d|\bm{P}_{CT}| \, d\cos\theta_C \, dM_{C}^2 \, d\phi_{R_C} \, d\eta_D \, d|\bm{P}_{DT}| \, d\cos\theta_D \, dM_{D}^2 \, d\phi_{R_D}} = \nonumber \\[0.3cm]
&\quad = \frac{|\bm{P}_{CT}| |\bm{P}_{DT}|}{8\pi^2} \, \sum_{a,b} \int \frac{dx_A dx_B dz_{hhC} dz_{hhD}}{z_{hhC}^2 z_{hhD}^2} \, f_1^a (x_A)\, f_1^b (x_B) \, x_B \delta (x_B - \bar{x}_B) \nonumber \\
&\quad \times z_{hhC} \delta (z_{hhC} - \bar{z}_{hhC}) \, \frac{z_{hhC} z_{hhD}^2}{|\bm{P}_{CT}| |\bm{P}_{DT}|} \, \delta (z_{hhD} - \bar{z}_{hhD}) \nonumber \\
&\quad \times \Bigg\{ \sum_{c,d} \Bigg[ \, \frac{d\hat{\sigma}_{ab\to cd}}{d\hat{t}} \, D_1^c (z_{hhC}, \cos\theta_C, M_{C}^2) \, D_1^d (z_{hhD}, \cos\theta_D, M_{D}^2) \nonumber \\
&\hspace{1.5cm} + \cos (\phi_{R_C} - \phi_{R_D}) \, \frac{d\Delta\hat{\sigma}_{ab \to c^\uparrow d^\uparrow}}{d\hat{t}} \, \frac{|\bm{R}_C|}{M_{C}} \, \sin \theta_C \, H_1^{\sphericalangle \, c} (z_{hhC}, \cos\theta_C, M_{C}^2) \, \frac{|\bm{R}_D|}{M_{D}} \, \sin \theta_D \, H_1^{\sphericalangle \, d} (z_{hhD}, \cos\theta_D, M_{D}^2) \, \Bigg]\nonumber \\
& \quad + \cos (2\phi_{R_C} - 2\phi_{R_D}) \, \frac{d\Delta\hat{\sigma}_{ab \to g^\uparrow g^\uparrow}}{d\hat{t}} \, \frac{|\bm{R}_C|^2}{M_{C}^2} \, \sin^2 \theta_C \,H_1^{\sphericalangle g\,} (z_{hhC}, \cos\theta_C, M_{C}^2) \,  \frac{|\bm{R}_D|^2}{M_{D}^2} \, \sin^2 \theta_D \, \,H_1^{\sphericalangle g\,} (z_{hhD}, \cos\theta_D, M_{D}^2) \Bigg\}  \; , 
\label{e:pp2x2h_PRD}
\end{align}
where the momenta and the angles of the second hadron pair are defined in complete analogy with the first pair by replacing the labels $c,C$ with $d,D$. The additional delta functions are due to momentum conservation in the elementary $ab \to cd$ process both in the longitudinal direction of the $\hat{z}$ axis, identified with $\bm{P}_A$, and in the transverse plane. 

In collinear kinematics, the conservation in the transverse plane is trivially $\bm{P}_{CT}/z_{hhC} = - \bm{P}_{DT}/z_{hhD}$. This implies that the above cross section is integrated in the azimuthal angles of $\bm{P}_{CT}$ and $\bm{P}_{DT}$ with the condition $\phi_C = \phi_D + \pi$ and that the moduli are constrained by~\cite{Bacchetta:2005rm} 
\begin{align}
\delta \left( \frac{|\bm{P}_{CT}|}{z_{hhC}} - \frac{|\bm{P}_{DT}|}{z_{hhD}} \right) &= \frac{z_{hhD}^2 z_{hhC}}{|\bm{P}_{CT}| |\bm{P}_{DT}|} \, \delta ( z_{hhD} - \bar{z}_{hhD} ) \; , \qquad 
\bar{z}_{hhD} = \frac{|\bm{P}_{DT}|}{\sqrt{s}} \, \frac{e^{\eta_C} + e^{\eta_D}}{x_A} \; .
\label{e:deltaPT}
\end{align}
The conservation along the $\hat{z}$ axis in the c.m. frame of the annihilation reads $x_A P_{Az} - x_B P_{Bz} = P_{Cz}/z_{hhC} + P_{Dz}/z_{hhD}$. Using the previous delta function, after some manipulation it can be rewritten as
\begin{align}
\delta \left( \eta_C + \eta_D + \log \frac{x_B}{x_A} \right) &= x_B \delta ( x_B - \bar{x}_B ) \; , \qquad 
\bar{x}_B = x_A e^{-\eta_C} e^{-\eta_D} \; . 
\label{e:deltaPL}
\end{align}
Finally, the third delta function is the analogue of Eq.~\eqref{e:ab2cd_mom}. Because of Eq.~\eqref{e:deltaPL}, it can be rewritten as
\begin{align}
\hat{s} \delta (\hat{s} + \hat{t} + \hat{u}) &= z_{hhC} \delta ( z_{hhC} - \bar{z}_{hhC}) \; , \qquad 
\bar{z}_{hhC} = \frac{|\bm{P}_{CT}|}{\sqrt{s}} \, \frac{x_A \, e^{-\eta_C} + x_B \, e^{\eta_C}}{x_A \, x_B} = 
\frac{|\bm{P}_{CT}|}{\sqrt{s}} \, \frac{e^{\eta_C} + e^{\eta_D}}{x_A} \; .
\label{e:zcbar}
\end{align}

In Eq.~\eqref{e:pp2x2h_PRD}, the elementary cross sections $d\Delta\hat{\sigma}_{ab \to c^\uparrow d^\uparrow}$ involve only quarks for the final partons $c, d$, while $d\Delta\hat{\sigma}_{ab \to g^\uparrow g^\uparrow}$ contain only final gluons linearly polarized in the transverse plane. Hence, the $H_1^{\sphericalangle g}$ function describes the fragmentation of such linearly polarized gluons into pairs of unpolarized hadrons. For both cases of final polarized quarks and gluons, all nonvanishing combinations are listed in the Appendix of Ref.~\cite{Bacchetta:2004it}. 

By introducing the same transformation of variables used for the inclusive production of a single hadron pair, the cross section of Eq.~\eqref{e:pp2x2h_PRD} can be rewritten as
\begin{align}
&\frac{d\sigma_{UU}}{d\eta_C \, d|\bm{P}_{CT}| \, d\zeta_C \, d\bm{R}_{C\perp} \, d\eta_D \, d|\bm{P}_{DT}| \, d\zeta_D \, d\bm{R}_{D\perp}} = \frac{|\bm{P}_{CT}| |\bm{P}_{DT}|}{8\pi^2} \, \sum_{a,b} \int \frac{dx_A dx_B dz_{hhC} dz_{hhD}}{z_{hhC}^2 z_{hhD}^2} \, f_1^a (x_A)\, f_1^b (x_B) \nonumber \\[0.2cm] 
&\quad \times \hat{s} \delta (\hat{s} + \hat{t} + \hat{u}) \, \delta \left( \frac{|\bm{P}_{CT}|}{z_{hhC}} - \frac{|\bm{P}_{DT}|}{z_{hhD}} \right)  \, \delta \left( x_A P_{Az} - x_B P_{Bz} - \frac{P_{Cz}}{z_{hhC}} - \frac{P_{Dz}}{z_{hhD}} \right)  \nonumber \\
&\quad \times \Bigg\{ \sum_{c,d} \Bigg[ \, \frac{d\hat{\sigma}_{ab\to cd}}{d\hat{t}} \, D_1^c (z_{hhC}, \zeta_C, \bm{R}_{C\perp}^2) \, D_1^d (z_{hhD}, \zeta_D, \bm{R}_{D\perp}^2) \nonumber \\
&\hspace{1.5cm} + \cos (\phi_{R_C} - \phi_{R_D}) \, \frac{d\Delta\hat{\sigma}_{ab \to c^\uparrow d^\uparrow}}{d\hat{t}} \, \frac{|\bm{R}_{C\perp}|}{M_{C}} \, H_1^{\sphericalangle \, c} (z_{hhC}, \zeta_C, \bm{R}_{C\perp}^2) \, \frac{|\bm{R}_{D\perp}|}{M_{D}} \, H_1^{\sphericalangle \, d} (z_{hhD}, \zeta_D, \bm{R}_{D\perp}^2) \, \Bigg] \nonumber \\
& \qquad + \cos (2\phi_{R_C} - 2\phi_{R_D}) \, \frac{d\Delta\hat{\sigma}_{ab \to g^\uparrow g^\uparrow}}{d\hat{t}} \, \frac{|\bm{R}_{C\perp}|^2}{M_{C}^2} \, \,H_1^{\sphericalangle g\,} (z_{hhC}, \zeta_C, \bm{R}_{C\perp}^2) \,  \frac{|\bm{R}_{D\perp}|^2}{M_{D}^2} \, \,H_1^{\sphericalangle g\,} (z_{hhD}, \zeta_D, \bm{R}_{D\perp}^2) \Bigg\}  \; . 
\label{e:pp2x2h_cmp}
\end{align}

%%%%%%%%%%%%%%%%%%%%%%%%%%%%%%%%%%%%%%%%%%%%%%%%%%%%%%
\section{Elementary hard cross section for $a+b \to c+d$}
\label{a:hard_xsec}

We compare the elementary cross section used in Refs.~\cite{Bacchetta:2004it,Kang:2017btw} and in the relevant literature for inclusive production of a hadronic final state in hadron-hadron collisions. 

In Refs.~\cite{D'Alesio:2004up,Anselmino:2004ky}, the cross section for the process $A+B \to C + X$ reads
\begin{equation}
\frac{E_C d\sigma}{d\bm{P}_C} = \sum_{abcd} \int \frac{dx_A dx_B dz_C}{\pi z_C^2} \, f_1^a (x_A) \, f_1^b (x_B) \, \frac{d\hat\sigma_{ab\to cd}}{d\hat{t}} \, \hat{s} \, \delta (\hat{s} + \hat{t} + \hat{u}) \, D_1(z_C) \; .
\label{e:To-Ca}
\end{equation}
The same structure of cross section can be obtained in Ref.~\cite{Aversa:1988vb} after making the transformation of variables $\hat{v} = 1 + \hat{t}/\hat{s}$ and $\hat{w} = - \hat{u} / (\hat{s} + \hat{t})$, but obtaining the above elementary cross section multiplied by $\hat{s}$. By labeling the $d\hat{\sigma}$ of Refs.~\cite{D'Alesio:2004up,Anselmino:2004ky} as $d\hat{\sigma}^{\mathrm{ToCa}}$ and the one of Ref.~\cite{Aversa:1988vb} as $d\hat{\sigma}^{\mathrm{ACGG}}$, we get 
\begin{equation}
\frac{d\hat{\sigma}^{\mathrm{ACGG}}_{ab\to cd}}{d\hat{t}} = \hat{s} \, \frac{d\hat{\sigma}^{\mathrm{ToCa}}_{ab\to cd}}{d\hat{t}} \; . 
\end{equation}

Equation~\eqref{e:To-Ca} can be further integrated in the angle of $\bm{P}_T$ and made differential in the pseudorapidity $\eta$ of the final hadron through the transformation $2\pi |\bm{P}_T| d|\bm{P}_T| dP_z / E \to 2\pi |\bm{P}_T| d|\bm{P}_T| d\eta$: 
\begin{equation}
\frac{d\sigma}{d\eta d|\bm{P}_T|} = 2|\bm{P}_T| \sum_{a,b,c,d} \int \frac{dx_A dx_B dz_C}{z_C^2} \, f_1^a (x_A) \, f_1^b (x_B) \, \frac{d\hat\sigma_{ab\to cd}^{\mathrm{ToCa}}}{d\hat{t}} \, \hat{s} \, \delta (\hat{s} + \hat{t} + \hat{u}) \, D_1(z_C) \; .
\label{e:To-Ca2}
\end{equation}
This expression is formally identical to the cross section for the $A + B \to (C_1 \, C_2) + X$ process of Eq.~\eqref{e:pp2h_PRD} after integrating over $d\cos\theta_C, \, d\phi_R$ and using Eq.~\eqref{e:ab2cd_mom} (apart for the dependence on the dihadron invariant mass $M_{hh}$ through the unpolarized DiFF $D_1$, which does not affect the argument about the elementary cross section). By labeling the $d\hat{\sigma}$ of Eq.~\eqref{e:pp2h_PRD} as $d\hat{\sigma}^{\mathrm{DiFF}}$, we get
\begin{equation}
\frac{d\hat{\sigma}^{\mathrm{ACGG}}_{ab\to cd}}{d\hat{t}} = \hat{s} \, \frac{d\hat{\sigma}^{\mathrm{ToCa}}_{ab\to cd}}{d\hat{t}} = \hat{s} \, \frac{d\hat{\sigma}^{\mathrm{DiFF}}_{ab\to cd}}{d\hat{t}} \; . 
\label{e:hardsigcomp}
\end{equation}
The above relation can be cross-checked by inspecting the elementary cross sections of various partonic channels in the Appendix of Refs.~\cite{Aversa:1988vb} and \cite{Bacchetta:2004it}, respectively. 

By recalling the relations~\eqref{e:hatMandelst} and~\eqref{e:ab2cd_mom}, Eq.~\eqref{e:h-in-jet_xsec} has the same structure as Eq.~\eqref{e:To-Ca2} (apart for the dependence on the variables $(z_h, \bm{j}_\perp)$ describing the jet substructure through the unpolarized jTMDFF ${\cal D}_1$, which does not affect the argument about the elementary cross section) provided that 
\begin{equation}
\frac{\pi z_{JC}}{\hat{s}} \, H^U_{ab\to cd}  = \frac{d\hat{\sigma}^{\mathrm{ToCa}}_{ab\to cd}}{d\hat{t}}  \; .
\label{e:hardsigcomp2}
\end{equation}
Because of Eq.~\eqref{e:hardsigcomp}, the above relation leads to Eq.~\eqref{e:hard_cmp}.

%Refs.~\cite{Kaufmann:2015hma,Kang:2017glf}

%Refer to Appendix to deduce this correspondence through the following steps:
%- write jet cross section  E dsigma / dvecPT  starting from DiFF one as dsigma / deta dPT dcosth dM2 dphiR dphiS 
% [ Sum_(h1,h2) \int dzc zc delta(zc-1) = 1]
%- do the same starting from 1h-emission as in TO-CA  PRD 2004 or 2005 => same elementary cross section
%- do the same using the jet cross section of Aversa et al.  (because are used by Kang) => correspondence on elementary cross sections 
%N.B. elementary cross section at LO is the same for 1h-emission quoted in the paper: 12->3  is the same of  12->j
%- do the same starting from Kang et al. in JHEP 2017 with Sum_h \int dzh dvecjperp zh D1 (zh, jperp2) = 1  => correspondence H with dsigma / dthat

%%%%%%%%%%%%%%%%%%%%%%%%%%%%%%%%%%%%%%%%%%%%%%%%%%%%%%

\bibliographystyle{apsrevM}
\bibliography{mybiblio}
%%%%%%%%%%%%%%%%%%%%%%%%%%%%%%%%%%%%%%%%%%%%%%%%%%%%%%

\end{document}